\pdfoutput=1

\documentclass[iop]{emulateapj}
\usepackage{natbib}
\usepackage{graphicx}
\usepackage{epstopdf}
\usepackage{lineno}
\usepackage{footnote}
\usepackage{epsfig}
\usepackage{graphics}
\usepackage{amssymb}
\usepackage{amsmath}
\usepackage{mathtools}
\usepackage{enumerate}
\usepackage{pifont}
\usepackage{hyperref}
\usepackage{cleveref}

\shorttitle{Optical variability modeling of blazar candidates behind the MCs}
\shortauthors{\.{Z}ywucka et al.}

\usepackage[utf8]{inputenc}
\usepackage[T1]{fontenc}
\usepackage{hyperref}

\def\dg{\hbox{$^\circ$}}

\def\utw{\smash{\rlap{\lower5pt\hbox{$\sim$}}}}
\def\udtw{\smash{\rlap{\lower6pt\hbox{$\approx$}}}}

\begin{document}

\title{Optical variability modeling of newly identified blazar candidates behind Magellanic Clouds}

\author{Natalia \.{Z}ywucka$^{1,2}$, Mariusz Tarnopolski$^1$,  Markus B\"{o}ttcher$^2$, {\L}ukasz Stawarz$^{1}$, and Volodymyr Marchenko$^{1}$}

\affil{$^1$Astronomical Observatory, Jagiellonian University, ul. Orla 171, 30-244 Kraków, Poland}
\affil{$^2$Centre of Space Research, North-West University, Potchefstroom, South Africa}
\email{n.zywucka@oa.uj.edu.pl,mariusz.tarnopolski@uj.edu.pl}

\begin{abstract}

We present an optical variability study of 44 newly identified blazar candidates behind the Magellanic Clouds, including 27 flat spectrum radio quasars (FSRQs) and 17 BL Lacertae objects (BL Lacs). All objects in the sample possess high photometric accuracy and irregularly sampled optical light curves (LCs) in I filter from the long-term monitoring conducted by the Optical Gravitational Lensing Experiment. We investigated the variability properties to look for blazar-like characteristics and to analyze the long-term behaviour. We analyzed the LCs with the Lomb-Scargle periodogram to construct power spectral densities (PSDs), found breaks for several objects, and linked them with accretion disk properties. In this way we constrained the black hole (BH) masses of 18 FSRQs to lie within the range $8.18\leq\log (M_{\rm BH}/M_\odot)\leq 10.84$, assuming a wide range of possible BH spins. By estimating the bolometric luminosities, we applied the fundamental plane of active galactic nuclei variability as an independent estimate, resulting in $8.4\leq\log (M_{\rm BH}/M_\odot)\leq 9.6$, with a mean error of 0.3. Many of the objects have very steep PSDs, with high frequency spectral index in the range $3-7$. An alternative attempt to classify the LCs was made using the Hurst exponent, $H$, and the $\mathcal{A}-\mathcal{T}$ plane. Two FSRQs and four BL Lacs yielded $H>0.5$, indicating presence of long-term memory in the underlying process governing the variability. Additionally, two FSRQs with exceptional PSDs, stand out also in the $\mathcal{A}-\mathcal{T}$ plane.

\end{abstract}


\section{Introduction}
The unification scheme of active galactic nuclei (AGNs) summarized by \cite{Urry95} defines blazars as type 0 radio-loud objects, $R =  F_{5\,\rm{GHz}}/F_{B}\geq 10$, where $F_{5\,\rm{GHz}}$ is the flux density at 5 GHz and $F_{\rm{B}}$ is the flux density in the B filter \citep{Kell89}, possessing unusual properties of optical emission lines and observed with their jets oriented at small angles \citep[$\lesssim 10$\dg; e.g.][]{Ange80,Falo14}. Based on the characteristics visible in the optical spectra, blazars are consistently divided into two groups: flat spectrum radio quasars (FSRQs), possessing prominent emission lines with the equivalent width of $>$5 \r{A}, and BL Lacertae objects (BL Lacs), having featureless continua or weak emission lines only. Blazars are characterized by non-thermal broad-band emission from radio up to $\gamma$-rays, high and variable polarization with the radio polarization degree at 1.4 GHz, PD$_{r,1.4}>$1\% \citep{Iler97}, flat radio spectra, $F_{\nu} \propto \nu^{-\alpha_r}$, with spectral index $\alpha_r<0.5$, and steep infrared to optical spectra, i.e. $0.5 \leq \alpha_o \leq 1.5$ \citep{Falo14}. Blazars show also rapid flux variability at all frequencies on different time scales from decades down to minutes. Variability patterns are generally divided into long-term variability continuing for years and decades \citep[e.g.][]{nilss18} and short-term variability with time scales from days up to months \citep[e.g.][]{Rani10,Alek15} with an additional separation of intraday/intranight variability lasting a fraction of a day \citep[e.g.][]{Wagn95,Bach12}. Blazar variability is typically studied in different separate energy ranges, such as radio \citep[e.g.][]{Alle11,Park14,Rich14}, optical \citep[e.g.][]{Saga04,Gaur12,Ruan12}, and $\gamma$-rays \citep[e.g.][]{Gaur10,Sobo14}, as well as using the multiwavelength approach in individual sources \citep[e.g.][]{Hart96,Wagn96,chatt12,Rani13}.

Variability is a unique property of blazars which can be also used as a tool to distinguish them from other astrophysical objects. By selecting blazar candidates from extensive monitoring programs and sky surveys, such as Palomar-QUEST survey \citep[PQ;][]{Baue09} or Magellanic Quasar Survey \citep[MQS;][]{Kozl13}, one can identify new types of blazars, or at least increase a sample of blazars which are underrepresented in the most commonly discussed blazar lists and samples constructed/compiled based predominantly on spectral properties and flux levels. For instance, \cite{Baue09} analyzed data gathered by the PQ survey, listing the 3113 most variable objects in a 7200 deg$^2$ field. All of them vary by more than 0.4 mag, simultaneously in I and R filters, on time scales provided by the survey, which lasted for 3.5 years. Additional separation was made applying a span of 200 days to not include transients into the sample and all objects were visually checked to remove artifacts. Moreover, the sources were checked in terms of optical colors typical to common stellar types to exclude variable stars from the sample. 

Subsequently, \cite{Kozl13} looked for AGNs behind the Magellanic Clouds (MCs) by analyzing optical photometric data from MQS, which covers 42 deg$^{2}$ in the sky, i.e. 100\% of the Large Magellanic Cloud (LMC) and 70\% of the Small Magellanic Cloud (SMC). The MQS quasars were selected in a four-step procedure:
\begin{enumerate}
    \item The quasar candidates selection was based on the cross-match of mid-infrared and optical data \citep{Kozl09}. A sample of 4699 and 657 quasar candidates behind the LMC and SMC, respectively, was selected. 
    \item Optical light curves (LCs) of all selected quasar candidates were analyzed using two methods, i.e. fitting a damped random walk model \citep{Kozl10}, and employing the structure function \citep{Kozl16}. This allowed the authors to define the MQS sample containing over 1000 quasar candidates.
    \item All variable OGLE sources behind the MCs were cross-matched with the X-ray data. As a result, \citet{Kozl12} selected a sample of 205 objects.  
    \item 3014 objects selected with at least one of the aforementioned steps were observed spectroscopically to verify if they are quasars. Eventually, \citet{Kozl13} listed 756 sources in the MQS catalog, including 565 quasars behind the LMC and 193 quasars behind the SMC.   
\end{enumerate}
Since blazars are expected to be more variable than other AGNs, the MQS catalog constitutes an excellent sample to look for new FSRQ blazar candidates. In addition, we searched for the BL Lac candidates using a list of sources rejected based on spectroscopic observations, i.e. among object possessing featureless optical spectra.

In our previous work \citep{Zywu18}, we identified a sample of 44 blazar candidates, including 27 FSRQs and 17 BL Lacs, whereof only nine objects (six FSRQs and three BL Lacs) were considered as secure blazar candidates. All objects in the sample were selected based on their radio, mid-infrared, and optical properties.
\begin{itemize}
    \item The blazar candidates were selected with the cross-matching procedure of radio and optical data.
    \item The characteristic properties of blazars, i.e. radio and mid-infrared indices as well as the radio-loudness parameters, were verified. All selected blazar candidates are distant objects with redshifts ranging from 0.29 up to 3.32, optically faint with the I band magnitude between 17.66 and 21.27, and radio-loud with $R \in [12, 4450]$ in the case of the FSRQ candidates, and $R \in [171, 7020]$ for the BL Lac candidates.
    \item The fractional linear polarization was checked or measured. We were able to collect the radio polarimetry parameters for nine objects from the AT20G sky survey catalog \citep{Murp10} and by analyzing polarized flux density maps at 4.8 and 8.6 GHz for the LMC\footnote{http://www.atnf.csiro.au/research/lmc\_ctm/index.html} and SMC\footnote{http://www.atnf.csiro.au/research/smc\_ctm/index.html}.
\end{itemize}
  We did not find any associations with X-rays and high energy $\gamma$-rays, cross-matching the ROSAT All-Sky Survey Catalogue \citep{Voge99} and the \textit{Fermi} 2FGL catalog \citep{Nola12}. However, the \textit{Fermi}-Large Area Telescope detected two flaring activities from direction of the J0545$-$6846 BL Lac candidate and we are currently checking a possible coincidence of this object and the $\gamma$-ray transient. 

Here, we extend the analysis of our blazar candidates with modeling of optical LCs provided by the OGLE group. All objects were selected from the long-term, deep optical monitoring survey, therefore they constitute a sample of faint sources with irregularly sampled optical LCs. We investigate them to determine variability-based classification of the blazar candidates and to analyze long-term behavior.   
    
This paper is organized as follows. Section~\ref{ogle} characterizes the LCs. In Section~\ref{method} we describe the methodology used to analyze the variability of the sources: Lomb-Scargle periodogram (LSP), Hurst exponent, and the $\mathcal{A}-\mathcal{T}$ plane. Section~\ref{results} summarizes the results obtained for all blazar candidates, and highlights the most interesting objects. Section~\ref{discussion} is devoted to discussion, and Section~\ref{conclusions} gives concluding remarks.

\section{OGLE light curves}
\label{ogle}

The blazar candidates considered here were selected from the well-monitored OGLE-III phase of the OGLE experiment \citep{Udal08a,Udal08b} and observed in the I optical filter, using the 1.3 m Warsaw telescope located at the Las Campanas Observatory in Chile. All 44 sources possess the OGLE-III data. The majority of objects, i.e. 15 FSRQs and 25 BL Lacs, were additionally monitored within the OGLE-IV phase \citep{Udal15}, while four blazar candidates, i.e. three FSRQs and one BL Lac, have also data from the OGLE-II phase \citep{Udal97}. This gives in total $\sim$17 years long LCs for objects with merged OGLE-II, -III, and -IV data, $\sim$12 years long LCs with OGLE-III and -IV data, and $\sim$7 years long LCs for sources with only OGLE-III data. 

In this study, we strive to include as much data as possible, but after the visual inspection of the original LCs we modified them as follows:
\begin{itemize}
\item points with uncertainties $>$10\% in magnitude were removed from all data sets,
\item obvious outliers in all LCs were removed manually; these may be errors in the data,
\item we discarded the OGLE-II data of the BL Lac candidate J0521-6959 due to the high noise level.
\end{itemize}

\begin{figure*}
\includegraphics[width=\textwidth]{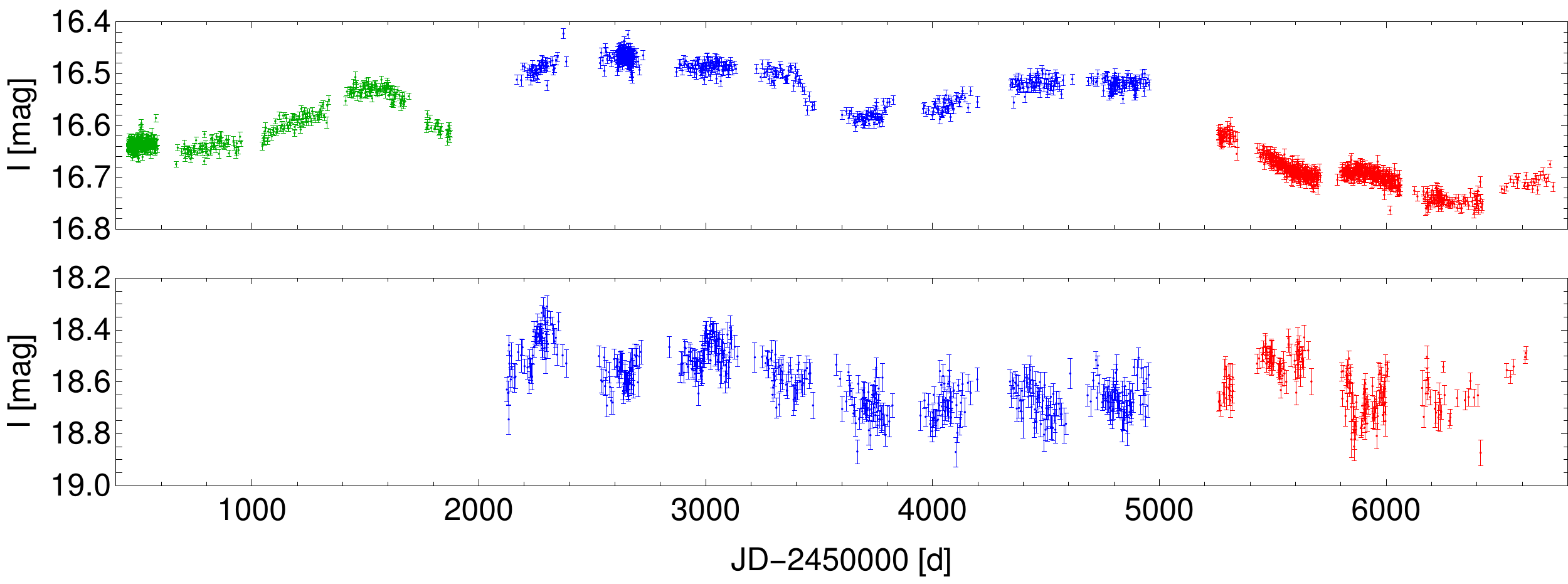}
\caption{Exemplary I band LCs of blazar candidates from our sample: the J0532--6931 FSRQ candidate (top panel) and the J0518-6755 BL Lac candidate (bottom panel). The OGLE-II data are shown with green color, OGLE-III with blue color, and OGLE-IV with red color.}
\label{plot_LCs}
\end{figure*}

 The exemplary LCs are shown in Fig.~\ref{plot_LCs}. By rejecting points with uncertainties $>$10\%, we lose only $\approx 1\%$ of data and it does not significantly affect the results. All LCs are sampled irregularly with short, medium, and long time intervals between observations. Most of the objects were observed with a time step $\Delta t\approx 1\,{\rm d}$, with gaps lasting up to a few days. These gaps were caused by bad weather conditions on the site. The medium time intervals are regular breaks in observations within the same OGLE phase, lasting between 3 and 5 months. During these times the MCs were too low to perform observations. After the OGLE-III phase, a technical upgrade of the telescope was performed, which resulted in a break between the OGLE-III and -IV phases, reaching 10--15 months. 

\section{Methodology} \label{method}

\subsection{Lomb-Scargle periodogram}
\label{sect3.1}

The LSP \citep{lomb,scargle82,press89,vanderplas18} is a way for constructing a power spectral density (PSD) for arbitrarily spaced data (unevenly sampled time series). For an LC with $N$ observations $x_k$ at times $t_k$, it is computed as
\begin{equation}
\begin{split}
P_{LS}(\omega) = \frac{1}{2\sigma^2} &\left[ \frac{\left(\sum\limits_{k=1}^N (x_k-\bar{x}) \cos[\omega(t_k-\tau)]\right)^2}{\sum\limits_{k=1}^N \cos^2[\omega(t_k-\tau)]} \right. \\
+& \left. \frac{\left(\sum\limits_{k=1}^N (x_k-\bar{x}) \sin[\omega(t_k-\tau)]\right)^2}{\sum\limits_{k=1}^N \sin^2[\omega(t_k-\tau)]}  \right],
\end{split}
\label{}
\end{equation}
where $\omega = 2\pi f$ is the angular frequency, $\tau\equiv\tau(\omega)$ is defined as
\begin{equation}
\tau(\omega)=\frac{1}{2\omega} \arctan \left[ \frac{\sum\limits_{k=1}^N\sin(2\omega t_k)}{\sum\limits_{k=1}^N\cos(2\omega t_k)} \right],
\label{}
\end{equation}
$\bar{x}$ and $\sigma^2$ are the sample mean and variance.

The lower limit for the sampled frequencies is $f_{\rm min} = 1/(t_{\rm max}-t_{\rm min})$, corresponding to the length of the time series. The upper limit, $f_{\rm max}$, would be the Nyquist frequency, the same as in case of the Fourier spectrum, if the data were uniformly sampled. For unevenly spaced data, the common choices for a pseudo-Nyquist frequency are somewhat arbitrary \citep{vanderplas18}. The upper limit herein is set to correspond to variations on time scales of 1 day, i.e. the maximal frequency is $f_{\rm max}=1\,{\rm day}^{-1}$. This is motivated by the fact that the most common time step between consecutive observations is scattered around one day. The sampling during each observing night was very nonuniform, and the data uncertainties obscure any short-term variability, hence the study of intranight variability is beyond the scope of this work. The total number of sampling frequencies is set to be $N_P = n_0 \frac{f_{\rm max}}{f_{\rm min}}$, with $n_0 = 10$ employed hereinafter. The Poisson noise level, coming from the statistical noise due to uncertainties in the measurements, $\Delta x_k$, is given by
\begin{equation}
P_{\rm Poisson} = \frac{1}{2\sigma^2} \sum\limits_{k=1}^N \Delta x_k^2.
\label{poiss}
\end{equation}

\subsection{Fitting}
\label{sect3.2}

To fit a PSD model in log-log space, one needs to take into account that the evenly spaced frequencies $f$ are no longer uniformly spaced when logarithmized, i.e. their density is greatly increased at higher values, where the Poisson noise can be expected to be significant. A straightforward least squares fitting would then rely mostly on points clustered in one region of the $\log f$ values, i.e. at high frequencies. To circumvent this problem, binning is applied. The values $\log f$ of the raw LSP are binned into equal-width bins, with at least six points in a bin, and the representative frequencies are computed as the geometric mean in each bin. The PSD value in a bin is taken as the arithmetic mean of the logarithms of the respective PSD values in that bin \citep{papadakis93,isobe15}.

Fits of different models are compared using the small sample Akaike Information Criterion ($AIC_c$) given by
\begin{equation}
AIC_c=2p-2\mathcal{L}+\frac{2(p+1)(p+2)}{N-p-2}
\label{}
\end{equation}
where $\mathcal{L}$ is the log-likelihood, $p$ is the number of parameters, and $N$ is the number of fitted points \citep{akaike74,hurvich89,burnham04}. For a regression problem,
\begin{equation}
\mathcal{L} = -\frac{1}{2}N\ln\frac{RSS}{N},
\label{}
\end{equation}
where $RSS$ is the residual sum of squares; $p$ is an implicit variable in the log-likelihood. A preferred model is one that minimizes $AIC_c$.

What is essential in assessing the relative goodness of a fit in the $AIC_c$ method is the difference, $\Delta_i=AIC_{c,i}-AIC_{c,\rm min}$, between the $AIC_c$ of the $i$-th model and the one with the minimal value, $AIC_{c,\rm min}$. If $\Delta_i<2$, then there is substantial support for the $i$-th model (or the evidence against it is worth only a bare mention), and the proposition that it is an adequate description is highly probable. In other words, both models are equally good, and one can not decide which one is better based only on the information criterion. A possible decision might be to choose the simpler model. Subsequently, if $2<\Delta_i<4$, then there is strong support for the $i$-th model. When $4<\Delta_i<7$, there is considerably less support, and models with $\Delta_i>10$ essentially exhibit no support.

\subsection{Hurst exponent}

The Hurst exponent $H$ \citep{hurst51,mandel68,katsev03,tarnopolski16d,knight17} measures the statistical self similarity of a time series $x(t)$. It is said that $x(t)$ is self similar (or self affine) if it satisfies
\begin{equation}
x(t)\stackrel{\textbf{\textrm{.}}}{=}\lambda^{-H}x(\lambda t),
\label{}
\end{equation}
where $\lambda>0$ and $\stackrel{\textbf{\textrm{.}}}{=}$ denotes equality in distribution. Self similarity is connected with long range dependence (memory) of a process via the autocorrelation function for lag $k$
\begin{equation}
\rho(k)=\frac{1}{2}\left[ (k+1)^{2H}-2k^{2H}+(k-1)^{2H} \right].
\label{}
\end{equation}
When $H>0.5$, $\rho(k)$ decays to zero as $k\rightarrow\infty$ so slowly that its accumulated sum does not converge (i.e., $\rho (k) \propto|k|^{-\delta}$, $0<\delta<1$), and $x(t)$ is then called a persistent process, i.e. the autocorrelations persist for a prolonged time. The persistency here relates to the higher probability of an increase (decrease) to be followed by another increase (decrease) in the short term, rather than alternating. Such a process is characterized by positive correlations at all lags. A process with $H<0.5$ is anti-persistent, also referred to as a mean-reverting series, i.e. having a tendency to quickly return to its long-term mean. Its autocorrelation $\rho (k)$ is summable. Both persistent and anti-persistent cases can be stationary and nonstationary (we consider weak stationarity herein).

\begin{figure*}
\includegraphics[width=\textwidth]{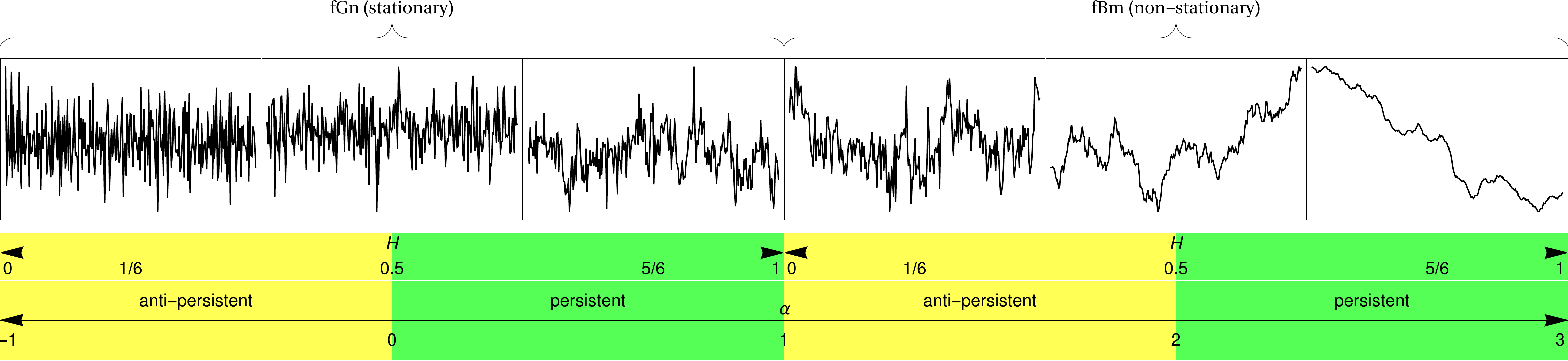}
\caption{The discontinuity of the Hurst exponent on the border between fGn and fBm. The high-$H$ fGn and low-$H$ fBm can be easily misidentified (figure based on \citealt{gilfriche18}).}
\label{fig_Hurst_disc}
\end{figure*}

The archetypal processes with long range dependence are the fractional Gaussian noise (fGn, a stationary process) and fractional Brownian motion (fBm, a nonstationary process with variance growing $\propto t^{2H}$ with time; the increments of an fBm constitute an fGn with the same $H$). There is a discontinuity of $H$ at the border between the two, where an fGn with $H\lesssim 1$ is very similar to an fBm with $H\gtrsim 0$, as illustrated in Fig.~\ref{fig_Hurst_disc}, and the two cases can be easily misidentified. For $H=0.5$, fGn and fBm reduce to white noise and Brownian motion, respectively. The properties of $H$ can be summarized as:
\begin{enumerate}
\item $0<H<1$,
\item $H=0.5$ for an uncorrelated process,
\item $H>0.5$ for a persistent (long-term memory, correlated) process,
\item $H<0.5$ for an anti-persistent (short-term memory, anti-correlated) process.
\end{enumerate}

For irregularly sampled data, the Hurst exponent can be obtained without any interpolation of the examined time series \citep{knight17} with the use of the lifting wavelet transform algorithm called {\it lifting one coefficient at a time} (LOCAAT). The algorithm aims at producing a set of wavelet-like coefficients, $\{d_{j_r}\}_r$, whose variance obeys the relation
\begin{equation}
\log_2 {\rm var}(d_{j_r})=\alpha\cdot j^*+{\rm const.},
\label{eqB}
\end{equation}
where $j^*$ is an equivalent of the wavelet scale, constructed for a set of $j_r$ coefficients. Eventually, the value of $\alpha$ is obtained via a linear regression of Eq.~(\ref{eqB}), and is linearly related with $H$ via $H=\frac{\alpha-1}{2}$ when $\alpha\in (1,3)$, and $H=\frac{\alpha+1}{2}$ when $\alpha\in (-1,1)$ (two most common instances, related to fBm- and fGn-like signals, respectively).\footnote{See also \citep{veitch99,tarnopolski15c,tarnopolski16d,knight17} and references therein for additional details on the Hurst exponent. For a detailed description of the LOCAAT, we refer the reader to \citet{knight17} and references therein.} Note the discontinuity of $H$ for the two ranges of $\alpha$ (see Fig.~\ref{fig_Hurst_disc}). We used the package \textsc{liftLRD}\footnote{\url{https://CRAN.R-project.org/package=liftLRD}} implemented in \textsc{R} to estimate $H$. The standard errors, obtained via bootstrapping, are returned by the package as well.

\subsection{The $\mathcal{A}-\mathcal{T}$ plane}
\label{meth_AT}

The Abbe value \citep{neumann41b,neumann41a,williams41,kendall1971,mowlavi14,tarnopolski16d} is defined as
\begin{equation}
\mathcal{A}=\frac{\frac{1}{N-1}\sum\limits_{k=1}^{N-1}(x_{k+1}-x_k)^2}{\frac{2}{N}\sum\limits_{k=1}^N (x_k-\bar{x})^2}.
\label{}
\end{equation}
It quantifies the smoothness of a time series by comparing the sum of the squared differences between two successive measurements with the standard deviation of the time series. It decreases to zero for time series displaying a high degree of smoothness, while the normalization factor ensures that $\mathcal{A}$ tends to unity for a purely noisy time series (more precisely, for a white noise process).

Three consecutive data points, $x_{k-1},x_k,x_{k+1}$, can be arranged in six ways; in four of them, they will create a peak or a valley, i.e. a turning point \citep{kendall73,brockwell96}. The probability of finding a turning point in such a subset is therefore $2/3$, and the expected value for a random data set is $\mu_T=\frac{2}{3}(N-2)$ --- the first and last ones cannot form turning points. Let $T$ denote the number of turning points in a time series, and $\mathcal{T}=T/N$ be their frequency relative to the number of observations. $\mathcal{T}$ is asymptotically equal to $2/3$ for a purely random time series (Gaussian noise). A time series with $\mathcal{T}>2/3$ (i.e., with raggedness exceeding that of a white noise) will be deemed more noisy than white noise. Similarly, a time series with $\mathcal{T}<2/3$ will be ragged less than Gaussian noise. The maximal value of $\mathcal{T}$ asymptotically approaches unity for a strictly alternating time series.
\begin{figure}
\centering
\includegraphics[width=\columnwidth]{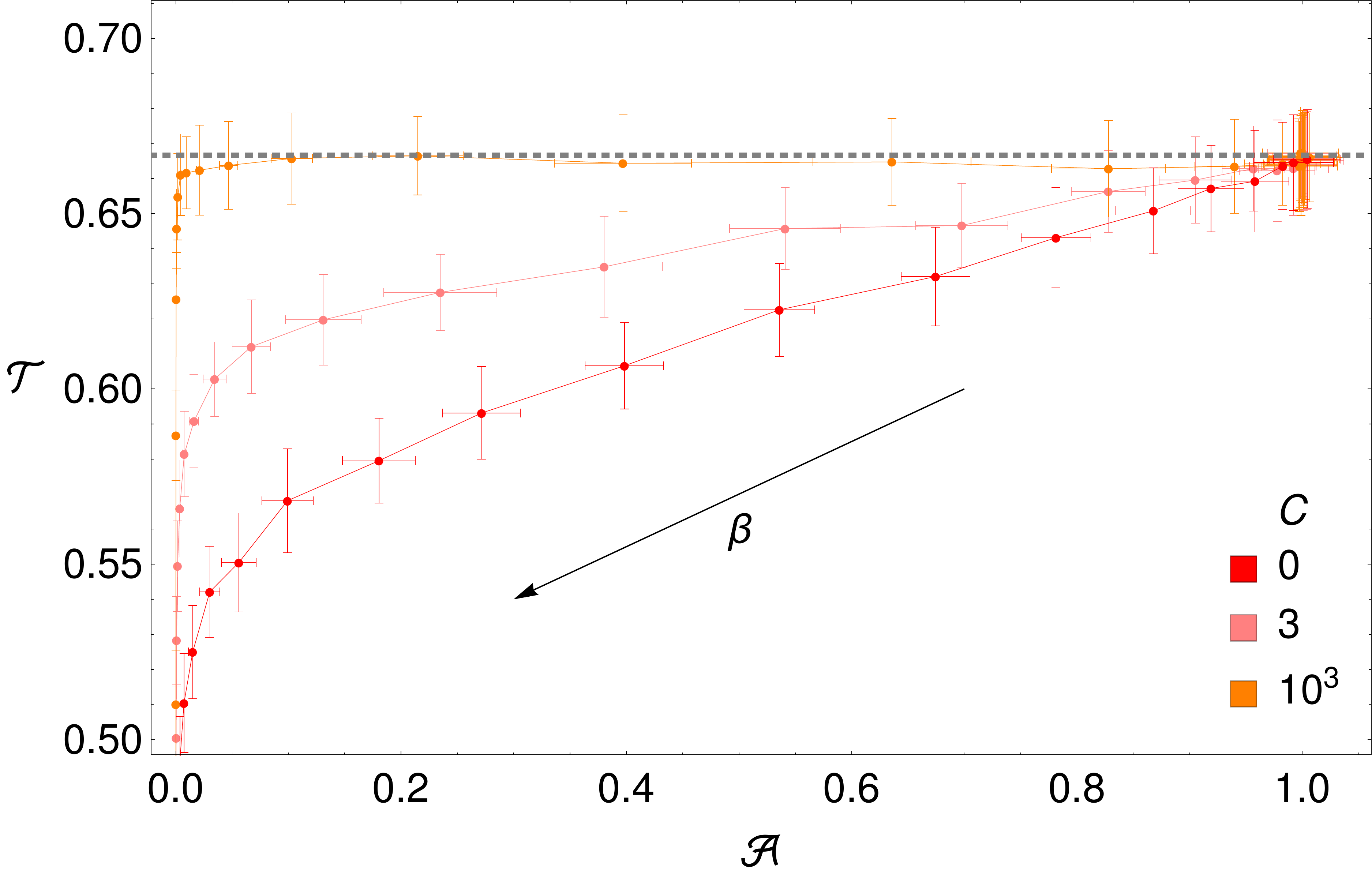}
\caption{Locations in the $\mathcal{A-T}$ plane of the PL plus Poisson noise PSD of the form $P(f)\propto 1/f^\beta+C$, with $\beta\in\{0,0.1,\ldots,3\}$ and where $C$ is the Poisson noise. For each PSD, 100 realizations of the time series were generated, and the displayed points are their mean locations. The error bars depict the standard deviation of $\mathcal{A}$ and $\mathcal{T}$ over these 100 realizations. The case $\beta=0$ is a pure white noise, with $(\mathcal{A},\mathcal{T})=(1,2/3)$. The generic PL case ($C=0$) is the lowest curve (red); with an increasing level of $C$ the curves are raised and shortened, as the white noise component starts to dominate over the PL part. The horizontal gray dashed line denotes $\mathcal{T}=2/3$ for Gaussian processeses.}
\label{fig_AT}
\end{figure}

The $\mathcal{A-T}$ plane \citep{tarnopolski16d} was initially introduced to provide a fast and simple estimate of the Hurst exponent. It is able to differentiate between different types of colored noise, $P(f)\propto 1/f^\beta$, characterized by different values of $\beta$. In Figure~\ref{fig_AT} we show the locations in the $\mathcal{A-T}$ plane of colored noise plus Poisson noise spectra of the form $P(f)\propto 1/f^\beta+C$, where the term $C$ is introduced to account for the uncertainties in the measurents, manifesting through the Poisson noise level from Eq.~(\ref{poiss}).

\section{Results} \label{results}

\begin{deluxetable*}{cllccccc}
\tabletypesize{\footnotesize}
\tablecolumns{8}
\tablewidth{0pt}
\tablecaption{Outcomes of the PSD modeling of newly identified FSRQ and BL Lac type blazar candidates. \label{blazarlist}}
\tablehead{
\multicolumn{1}{c}{Number} & \multicolumn{1}{c}{Object} & \colhead{$\beta$} & \colhead{$\log f_0$} & \colhead{$\beta_1$} & \colhead{$\beta_2$} & \colhead{$T_{\rm break}$} & \colhead{Best}  \\
 & & & [1/d] & & & [d] & model \\
\tiny{(1)} & \multicolumn{1}{c}{\tiny{(2)}} & \multicolumn{1}{c}{\tiny{(3)}} & \tiny{(4)} & \tiny{(5)} & \tiny{(6)} & \tiny{(7)} & \tiny{(8)}  }
\startdata
\multicolumn{8}{c}{FSRQ type blazar candidates}\\\hline
1 & J0054$-$7248 & $1.21\pm 0.13$ & $-2.16\pm 0.43$ & --- & --- & --- & A \\
2 & J0114$-$7320$^{\star}$ & $1.45\pm 0.18$ & $-1.65\pm 0.42$ & $0.37\pm 0.40$ & $4.53\pm 1.56$ & $338\pm 127$ & B \\
3 & J0120$-$7334$^{\star}$ & $1.86\pm 0.15$ & $-1.90\pm 0.28$ & $1.28\pm 0.23$ & $5.00\pm 1.91$ & $229\pm 75$ & B \\
4 & J0122$-$7152 & $1.45\pm 0.17$ & $-1.76\pm 0.40$ & $1.01\pm 0.27$ & $5.65\pm 4.35$ & $155\pm 61$ & A, B \\
5 & J0442$-$6818$^{\star}$ & $1.75\pm 0.27$ & $-1.98\pm 0.54$ & $0.83\pm 0.31$ & $8.93\pm 5.20$ & $241\pm 51$ & B \\
&&&&&&\\
6 & J0445$-$6859 & $1.30\pm 0.29$ & $-2.13\pm 0.77$ & --- & --- & --- & A \\
7 & J0446$-$6758 & $1.60\pm 0.18$ & $-1.92\pm 0.39$ & --- & --- & --- & A \\
8 & J0455$-$6933 & $1.58\pm 0.28$ & $-1.90\pm 0.61$ & $0.30\pm 0.36$ & $6.83\pm 3.38$ & $250\pm 58$ & B \\
9 & J0459$-$6756 & $1.64\pm 0.27$ & $-1.92\pm 0.56$ & $0.88\pm 0.44$ & $6.56\pm 5.47$ & $246\pm 103$ & A, B \\
10 & J0510$-$6941 & $1.63\pm 0.13$ & $-1.73\pm 0.28$ & $0.96\pm 0.16$ & $5.75\pm 1.58$ & $218\pm 39$ & B \\
&&&&&&\\
11 & J0512$-$7105 & $0.14\pm 0.06^\flat$ & --- & --- & --- & --- & A \\
12 & J0512$-$6732$^{\star}$ & $1.00\pm 0.12$ & $-1.32\pm 0.38$ & $0.64\pm 0.09$ & $6.84\pm 3.26$ & $67\pm 10$ & B \\
13 & J0515$-$6756 & $1.40\pm 0.24$ & $-2.41\pm 0.67$ & --- & --- & --- & A \\
14 & J0517$-$6759 & $1.23\pm 0.21$ & $-1.79\pm 0.58$ & $0.70\pm 0.30$ & $6.66\pm 6.04$ & $164\pm 56$ & A, B \\
15 & J0527$-$7036 & $1.26\pm 0.13$ & $-1.28\pm 0.33$ & $1.06\pm 0.22$ & $3.82\pm 3.69$ & $48\pm 34$ & A$^\natural$ \\
&&&&&&\\ 
16 & J0528$-$6836 & $1.12\pm 0.17$ & $-1.21\pm 0.47$ & --- & --- & --- & A \\
17 & J0532$-$6931 & $1.29\pm 0.14$ & $-1.10\pm 0.34$ & $0.68\pm 0.21$ & $4.33\pm 1.76$ & $115\pm 45$ & B \\
18 & J0535$-$7037 & $1.11\pm 0.14$ & $-2.31\pm 0.50$ & --- & --- & --- & A \\
19 & J0541$-$6800 & $1.56\pm 0.15$ & $-1.76\pm 0.32$ & $0.71\pm 0.34$ & $3.35\pm 1.00$ & $319\pm 172$ & B \\
20 & J0541$-$6815 & $1.92\pm 0.12$ & $-1.64\pm 0.45$ & $1.45\pm 0.16$ & $5.87\pm 1.76$ & $177\pm 34$ & B \\
&&&&&&\\
21 & J0547$-$7207 & $1.37\pm 0.21$ & $-1.69\pm 0.51$ & $0.29\pm 0.37$ & $4.66\pm 2.00$ & $284\pm 106$ & B \\
22 & J0551$-$6916$^{\star}$ & $1.46\pm 0.22$ & $-1.91\pm 0.54$ & $0.75\pm 0.31$ & $7.36\pm 5.21$ & $225\pm 64$ & B \\
23 & J0551$-$6843$^{\star}$ & $1.48\pm 0.17$ & $-1.72\pm 0.40$ & --- & --- & --- & A \\
24 & J0552$-$6850 & $1.62\pm 0.14$ & $-1.73\pm 0.30$ & $-0.70\pm 0.76$ & $2.36\pm 0.28$ & $1201\pm 417$ & B \\
25 & J0557$-$6944 & $1.57\pm 0.24$ & $-2.02\pm 0.55$ & --- & --- & --- & A \\
&&&&&&\\
26 & J0559$-$6920 & $1.44\pm 0.19$ & $-1.90\pm 0.46$ & $0.79\pm 0.33$ & $5.51\pm 3.57$ & $248\pm 93$ & A, B \\
27 & J0602$-$6830 & $1.35\pm 0.13$ & $-1.54\pm 0.32$ & $0.30\pm 0.69$ & $2.25\pm 0.68$ & $538\pm 579$ & B \\\hline
\multicolumn{8}{c}{BL Lac type blazar candidates}\\\hline
1 & J0039$-$7356 & $1.61\pm 0.28$ & $-2.74\pm 0.74$ & --- & --- & --- & A \\
2 & J0111$-$7302$^{\star}$ & $1.76\pm 0.44$ & $-2.50\pm 0.99$ & --- & --- & --- & A \\
3 & J0123$-$7236 & $4.02\pm 1.23$ & $-3.13\pm 1.42$ & --- & --- & --- & A \\
4 & J0439$-$6832 & $0.98\pm 0.22$ & $-2.15\pm 0.84$ & --- & --- & --- & A \\
5 & J0441$-$6945 & $1.20\pm 0.16$ & $-2.32\pm 0.52$ & --- & --- & --- & A \\
&&&&&&\\
6 & J0444$-$6729 & $1.47\pm 0.21$ & $-2.05\pm 0.51$ & $1.00\pm 0.48$ & $4.42\pm 4.20$ & $193\pm 133$ & A$^\natural$ \\
7 & J0446$-$6718 & $3.50 \pm 1.13$ & $-3.25\pm 1.52$ & --- & --- & --- & A \\
8 & J0453$-$6949 & $2.64\pm 0.66$ & $-2.91\pm 1.09$ & --- & --- & --- & A \\
9 & J0457$-$6920 & $1.03\pm 0.18$ & $-1.95\pm 0.64$ & --- & --- & --- & A \\
10 & J0501$-$6653$^{\star}$ & $1.44\pm 0.20$ & $-1.98\pm 0.50$ & $0.98\pm 0.32$ & $6.95\pm 6.63$ & $217\pm 78$ & A, B \\
&&&&&&\\
11 & J0516$-$6803 & $-0.04\pm 0.05^\flat$ & --- & --- & --- & --- & A \\
12 & J0518$-$6755$^{\star}$ & $1.34\pm 0.15$ & $-1.73\pm 0.39$ & $0.84\pm 0.33$ & $3.75\pm 2.28$ & $182\pm 118$ & A, B \\
13 & J0521$-$6959 & $1.16\pm 0.24$ & $-1.31\pm 0.58$ & --- & --- & --- & A \\
14 & J0522$-$7135 & $1.16\pm 0.39$ & $-2.59\pm 1.37$ & --- & --- & --- & A \\
15 & J0538$-$7225 & $1.09\pm 0.16$ & $-1.60\pm 0.50$ & $0.42\pm 0.25$ & $5.17\pm 2.86$ & $183\pm 57$ & B \\
&&&&&&\\
16 & J0545$-$6846 & $0.99\pm 0.38$ & $-2.48\pm 1.51$ & --- & --- & --- & A \\
17 & J0553$-$6845 & $1.34\pm 0.19$ & $-2.12\pm 0.53$ & --- & --- & --- & A 
\enddata 
\tablecomments{$^{\star}$Strongly polarized sources with the average radio polarization degree at 4.8 GHz, PD$_{r,4.8}\sim$6.8\%. These sources are considered as secure blazar candidates by \citet{Zywu18}.\\ $^\flat$Obtained by fitting a pure PL. \\ $^\natural$With $2<\Delta_i<4$. \\Columns: (1) number of the source; (2) source designation; (3) PL index of model A; (4) critical frequency of model A; (5) low frequency index of model B; (6) high frequency index of model B; (7) break time scale of model B; (8) best model. }
\end{deluxetable*}

\subsection{LSP}
\label{results_lsp}

The following models are fitted to the LSP of each object:
\begin{enumerate}
    \item model A: a power law (PL) plus Poisson noise
    \begin{equation}
        P(f) = \frac{P_{\rm norm}}{f^\beta}+C;
        \label{eqA}
    \end{equation}
    \item model B: a smoothly broken PL (SBPL) plus Poisson noise \citep{mchardy2004,alston19a,alston19b}:
    \begin{equation}
        P(f) = \frac{P_{\rm norm}f^{-\beta_1}}{1 + \left( \frac{f}{f_{\rm break}} \right)^{\beta_2-\beta_1}} + C,
    \end{equation}
\end{enumerate}
where the parameter $C$ is an estimate of the Poisson noise level from Eq.~(\ref{poiss}), $f_{\rm break}$ is the break frequency, and $\beta_1,\,\beta_2$ are the low and high frequency indices, respectively.

Parameters of the fits are gathered in Table~\ref{blazarlist}, where $T_{\rm break} = 1/f_{\rm break}$. The uncertainties are the standard errors of the fit, and $\Delta T_{\rm break}$ comes from the law of error propagation. The best model is chosen based on $AIC_c$ values\footnote{The same conclusions were arrived at when $BIC=p\ln N-2\mathcal{L}$ was employed \citep{schwarz78,kass95}.}. When $\Delta_i<2$, both models are equally good. Only 10 FSRQs clearly yield model A as the better one (with J0512$-$7105 consistent with a nearly flat PSD\footnote{Fitting model A yielded $\beta=0.14\pm 1.57$, hence a flat PSD, but with a huge error. This is due to the degeneracy of model A when $\beta\rightarrow 0$: $P(f)\approxeq P_{\rm norm}+C=\rm{const.}$ Therefore, a pure PL was fitted to remove this degeneracy, and its result is displayed in Table~\ref{blazarlist}.\label{footnote_label}}); for 13 FSRQs model B is preferred. In case of BL Lacs, 14 are best described by model A, and only in one instance (J0538$-$7225) model B was pointed at.

\begin{figure}
\centering
\includegraphics[width=\columnwidth]{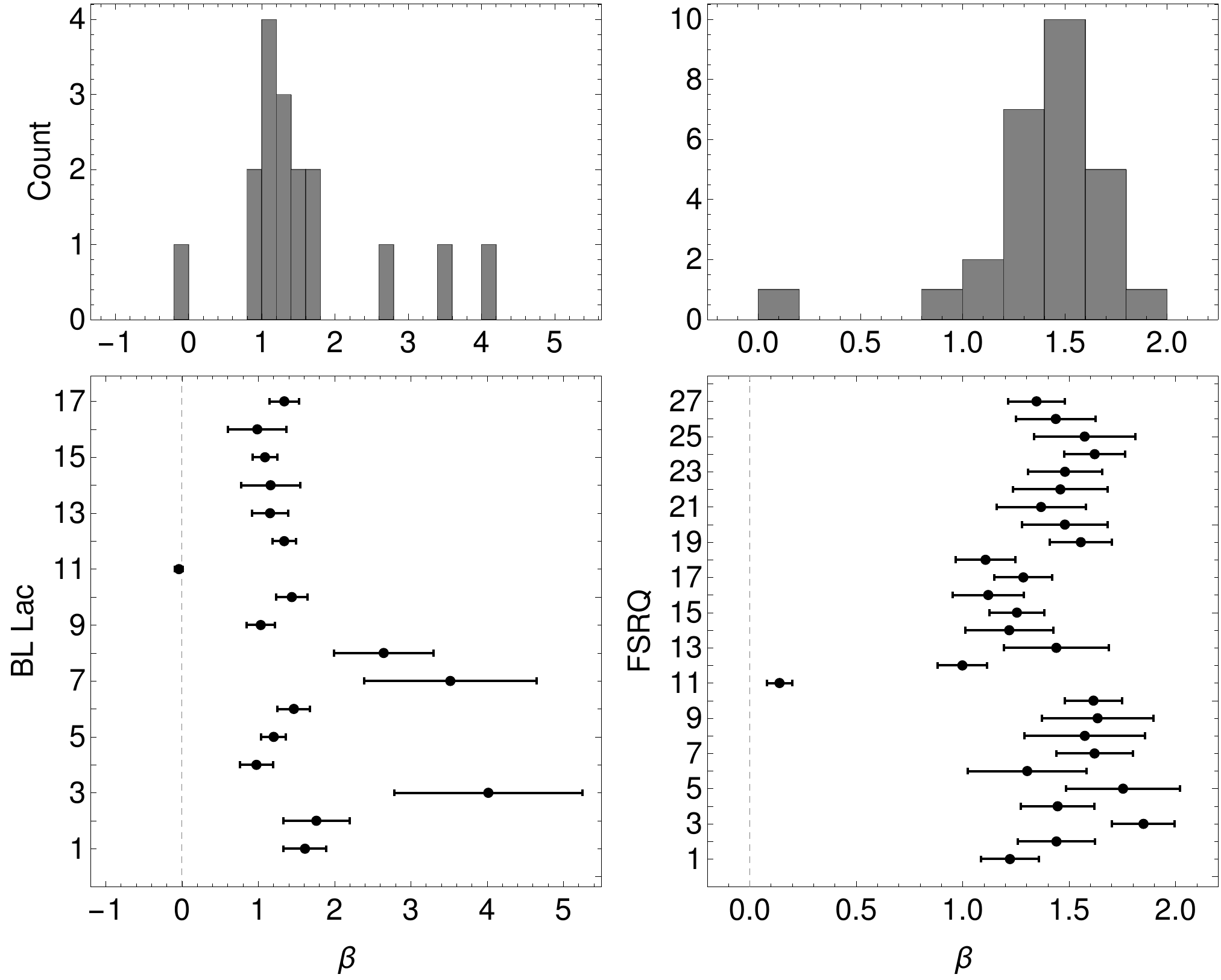}
\caption{Distributions of PL indices $\beta$ from model A fitted to the binned LSPs.}
\label{fig_ls}
\end{figure}

The distributions of the exponents $\beta$ from model A of all 44 objects are displayed in Fig.~\ref{fig_ls}. For FSRQs, the exponent $\beta$ mostly lies in the range $(1,2)$, with the mode at 1.5. BL Lacs are slightly flatter, spanning mostly the range $(1,1.8)$, with the mode at about 1.2; one object has a flat PSD\footnote{Model A yielded $\beta=0.00\pm 0.93$, hence a pure PL was fitted. See footnote \ref{footnote_label}.}. On the other hand, three BL Lacs have steeper PSDs, with $\beta \sim 3-4$. Fits of the models are shown in Figs.~\ref{fig_fits_FSRQ} and 
\ref{fig_fits_BLLAC} in Appendix~\ref{appC}.

To check the reliability of the fits, we performed Monte Carlo simulations described in Appendix~\ref{appA}. In short, one can expect the $\beta$ values from model A to be spread over a wide range for input $\beta\lesssim 2$, and slightly overestimated for $\beta\gtrsim 2$. Fitting to model B yields statistically reliable estimates, although with a non-negligible spread and greater uncertainties. Finally, in Appendix~\ref{appB} we tested for spurious detections of model B due to irregularly sampled LCs by imposing the same sampling as in the LCs of the examined FSRQs. We found that owing primarily to statistical fluctuations, model B should appear only 9 times, half of the actual case of 18 instances of model B being a plausible result. We can therefore conclude that a subpopulation of FSRQs, with PSDs given by model B, to be actually present in our sample is very likely.

\begin{figure}
\centering
\includegraphics[width=\columnwidth]{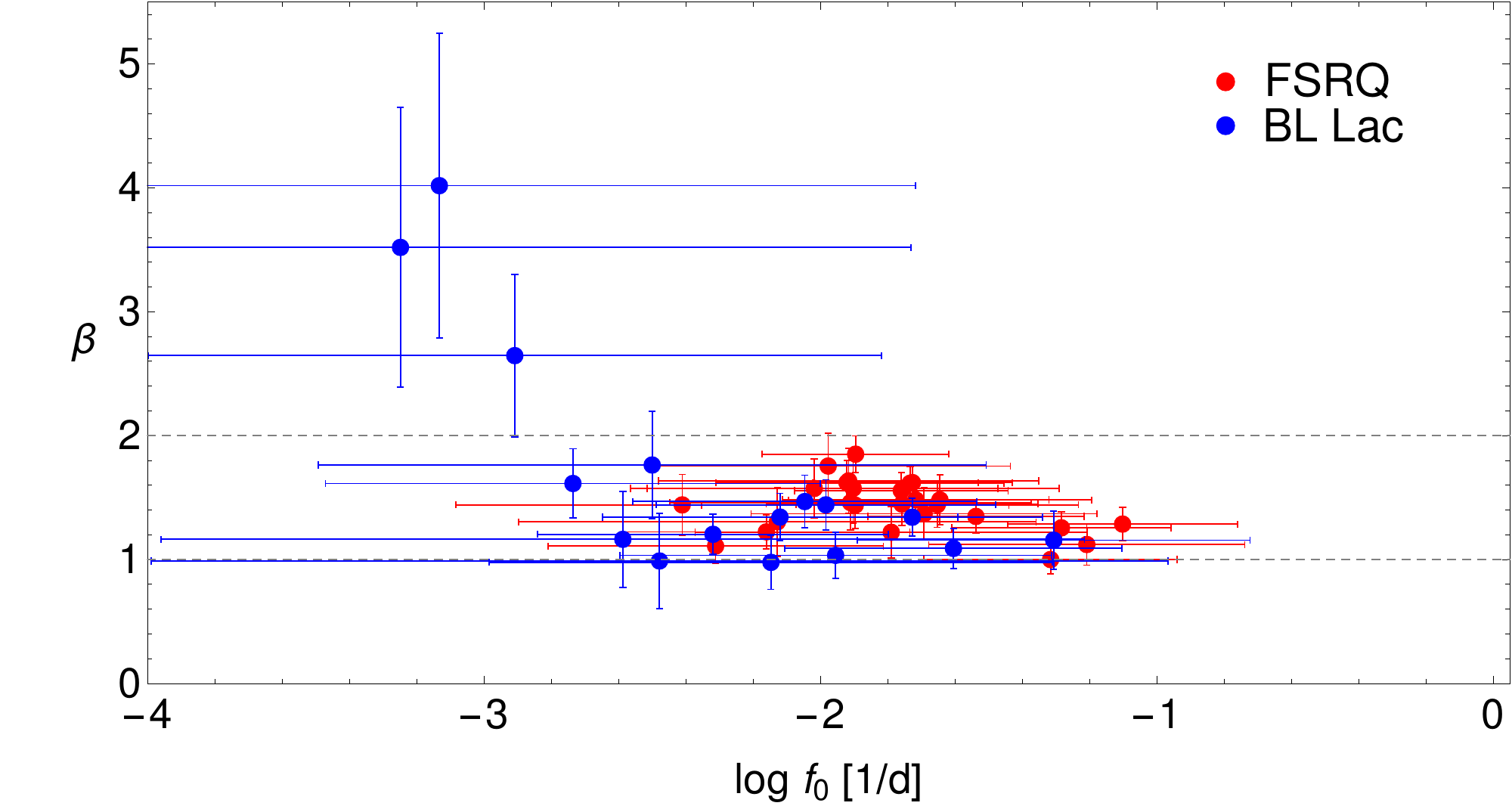}
\caption{Relation between the critical frequency $f_0$ and the PL index $\beta$ from model A.}
\label{fig_f0}
\end{figure}

Our BL Lac type objects are on average dimmer than FSRQs: $\bar{I} = 20.09\pm 0.18\,{\rm mag}$ and $\bar{I} = 19.00\pm0.20\,{\rm mag}$, respectively \citep{Zywu18}. This fact may be the reason why for a majority of BL Lacs model A is more adequate, while FSRQs require a more complex model B, i.e. the variability properties of BL Lacs could not be constrained on shorter time scales due to domination of Poisson noise at such scales. To test this, we calculated the critical frequency $f_0$ at which the Poisson noise has the same power as the PL component, $P_{\rm norm}/f_0^\beta = C \Rightarrow f_0 = (P_{\rm norm}/C)^{1/\beta}$. For all FSRQs, $\log f_0 \gtrsim -2.4$, and the corresponding $\beta\in (1,2)$. In case of BL Lacs, however, for the three objects with $\beta>2$, their $\log f_0 < -2.8$, and $I \gtrsim 20.5\,{\rm mag}$ (see Fig.~\ref{fig_f0} and \citealt{Zywu18}). This implies that their high $\beta$ values are artifacts of fitting model A to PSDs that are significantly dominated by Poisson noise over a wide range of time scales, including the longest ones, where only 1--2 points contribute to the PL part while fitting model A; hence are just statistical fluctuations. Note, however, that there are still three other BL Lacs that are even dimmer (J0039$-$7356, J0441$-$6945, and J0545$-$6846), but appear to be well described by model A. Overall, in all instances where model B was chosen as the better one, $\beta_2 \gtrsim 3$ (except for J0552$-$6850, which yields an unusually high $T_{\rm break} = 1201\pm 417$, and J0602$-$6830, whose $T_{\rm break} = 538\pm 579$ is not constrained well due to a huge uncertainty).

\begin{figure}
\centering
\includegraphics[width=\columnwidth]{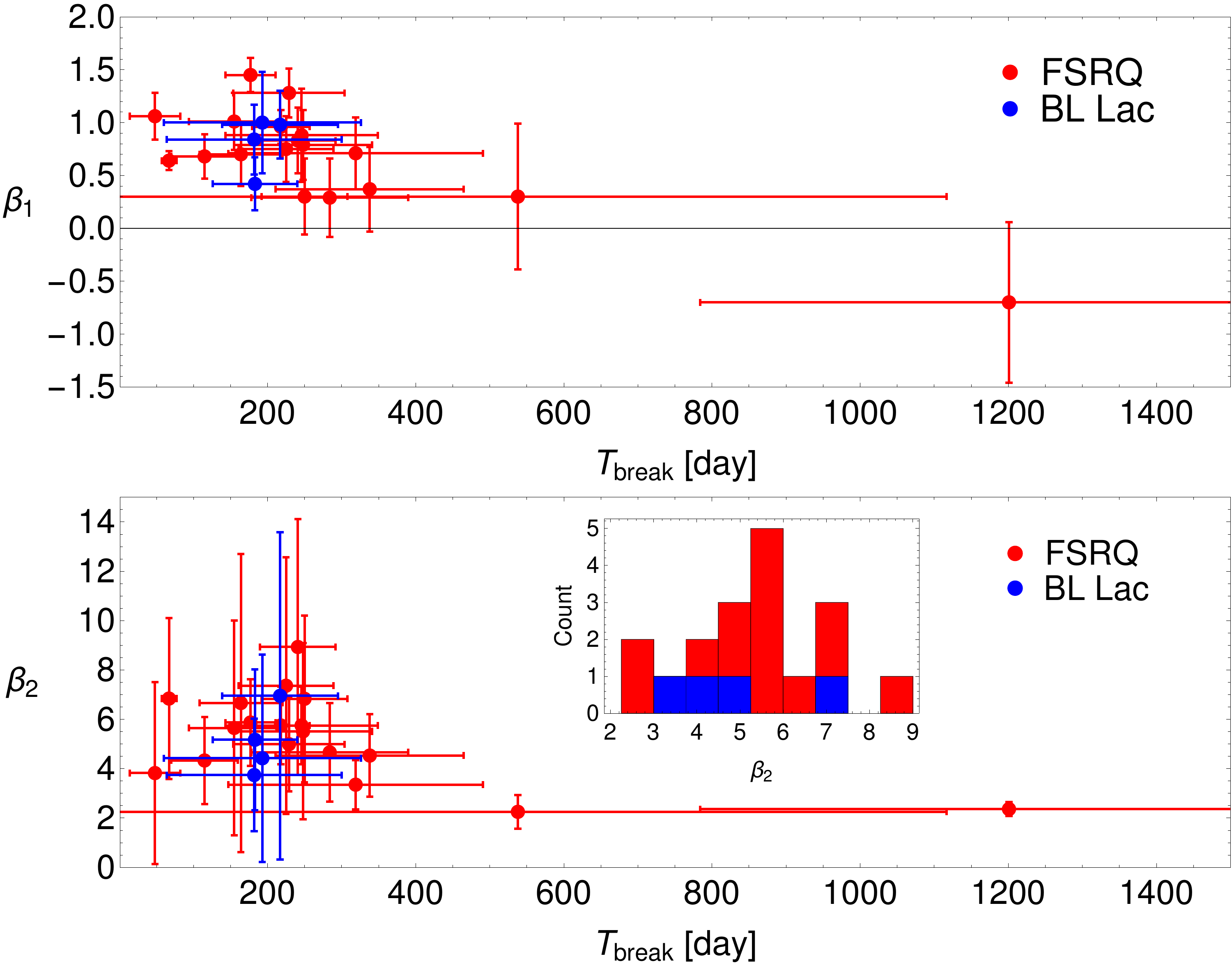}
\caption{Relations between the break time scale $T_{\rm break}$ and low-frequency PL index $\beta_1$ (upper panel) and high-frequency PL index $\beta_2$ (lower panel) from model B. The inset shows the distribution of $\beta_2$.}
\label{fig_break}
\end{figure}

Fig.~\ref{fig_break} presents the relations between $T_{\rm break}$ and the exponents $\beta_1$ and $\beta_2$ from model B. There are 18 FSRQs and 4 BL Lacs (including J0444$-$6729) that yielded reasonable fits. The Pearson correlation coefficient for the $\beta_2-T_{\rm break}$ relation is $r=-0.49$; after discarding the FSRQs with $T_{\rm break}>500\,{\rm d}$ and large errors, it reduces to $r=-0.01$. For the $\beta_1-T_{\rm break}$ relation, the respective correlation coefficients are $-0.78$ and $-0.33$. This indicates no significant correlation between the high-frequency PSD index $\beta_2$ and $T_{\rm break}$, and moderate correlation between the low-frequency index $\beta_1$ and $T_{\rm break}$. According to this, the relative power in the long time-scale variability increases with decreasing $T_{\rm break}$.  

\subsection{Hurst exponents}
\label{results_hurst}

\begin{figure}
\centering
\includegraphics[width=\columnwidth]{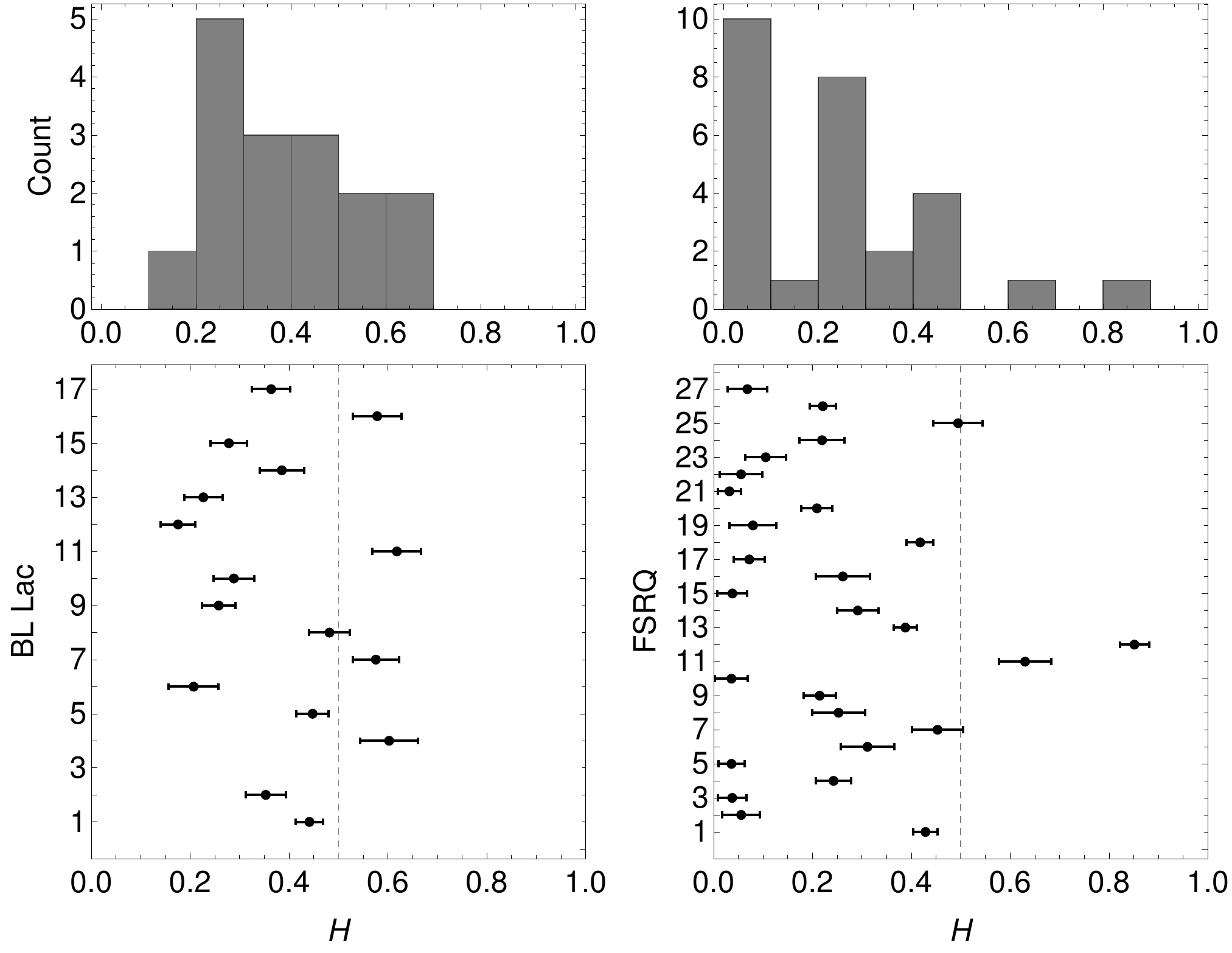}
\caption{Distributions of the Hurst exponents of examined BL Lac and FSRQ candidates.}
\label{fig_hurst}
\end{figure}

The PL form of a PSD is indicative of a self-affine stochastic process, characterized by the Hurst exponent $H$, underlying the observed variability. The $H$ values are listed in Table~\ref{blazarlist2}, and displayed graphically in Fig.~\ref{fig_hurst}. We find that most objects have $H\leq 0.5$ within errors, indicating short-term memory. Four BL Lacs (whose PSDs are best described by model A) and two FSRQs (J0512$-$7105 with a flat PSD, and J0512$-$6732, a secure blazar candidate with PSD given by model B, with an exceptionally short $T_{\rm break} = 67\pm 10$) yield $H>0.5$, implying long-term memory. Very few objects are characterized by $H\approx 0.5$, so the modeled stochastic process is not necessarily uncorrelated. There is also a number of FSRQs and a few BL Lacs with $H\lesssim 0.2$, i.e. close to the discontinuity value on the border between fGn- and fBm-like processes (see Fig.~\ref{fig_Hurst_disc}), hence their $H$ estimates are uncertain. In general, the autocorrelation functions drop to zero after time scales comparable to $T_{\rm break}$, above which the system becomes decorrelated \citep{caplar19}. This strongly suggests that models admitting long range dependence \citep{tsai05,tsai09,feigelson18} should be considered as candidates for the underlying stochastic processes governing the observed variability of blazar LCs.

\begin{deluxetable*}{clccc}
\tabletypesize{\footnotesize}
\tablecolumns{5}
\tablewidth{0pt}
\tablecaption{Hurst exponents and $(\mathcal{A},\mathcal{T})$ locations of the newly identified FSRQ and BL Lac type blazar candidates. \label{blazarlist2}}
\tablehead{
\multicolumn{1}{c}{Number} & \multicolumn{1}{c}{Object} & \colhead{$H$} & \colhead{$\mathcal{A}$} & \colhead{$\mathcal{T}$} \\
\tiny{(1)} & \multicolumn{1}{c}{\tiny{(2)}} & \multicolumn{1}{c}{\tiny{(3)}} & \tiny{(4)} & \tiny{(5)} }
\startdata
\multicolumn{5}{c}{FSRQ type blazar candidates}\\\hline
1 & J0054$-$7248 & $0.42\pm 0.02$ & $0.83\pm 0.02$ & $0.670\pm 0.011$ \\
2 & J0114$-$7320$^{\star}$ & $0.06\pm 0.04$ & $0.033\pm 0.002$ & $0.653\pm 0.015$ \\
3 & J0120$-$7334$^{\star}$ & $0.04\pm 0.03$ & $0.027\pm 0.002$ & $0.653\pm 0.012$ \\
4 & J0122$-$7152 & $0.24\pm 0.04$ & $0.24\pm 0.02$ & $0.643\pm 0.012$ \\
5 & J0442$-$6818$^{\star}$ & $0.04\pm 0.03$ & $0.022\pm 0.002$ & $0.664\pm 0.014$ \\
&&&&\\
6 & J0445$-$6859 & $0.31\pm 0.05$ & $0.59\pm 0.03$ & $0.667\pm 0.017$ \\
7 & J0446$-$6758 & $0.45\pm 0.05$ & $0.30\pm 0.02$ & $0.652\pm 0.011$ \\
8 & J0455$-$6933 & $0.25\pm 0.05$ & $0.22\pm 0.02$ & $0.671\pm 0.019$ \\
9 & J0459$-$6756 & $0.21\pm 0.03$ & $0.17\pm 0.01$ & $0.668\pm 0.012$ \\ 
10 & J0510$-$6941 & $0.04\pm 0.03$ & $0.20\pm 0.01$ & $0.637\pm 0.013$ \\
&&&&\\
11 & J0512$-$7105 & $0.63\pm 0.05$ & $0.90\pm 0.05$ & $0.717\pm 0.019$ \\
12 & J0512$-$6732$^{\star}$ & $0.85\pm 0.03$ & $0.26\pm 0.02$ & $0.640\pm 0.014$ \\
13 & J0515$-$6756 & $0.38\pm 0.02$ & $0.82\pm 0.03$ & $0.670\pm 0.012$ \\
14 & J0517$-$6759 & $0.29\pm 0.04$ & $0.53\pm 0.03$ & $0.663\pm 0.013$ \\ 
15 & J0527$-$7036 & $0.04\pm 0.03$ & $0.07\pm 0.01$ & $0.655\pm 0.013$ \\
&&&&\\ 
16 & J0528$-$6836 & $0.26\pm 0.05$ & $0.19\pm 0.01$ & $0.651\pm 0.011$ \\
17 & J0532$-$6931 & $0.07\pm 0.03$ & $0.013\pm 0.001$ & $0.626\pm 0.009$ \\
18 & J0535$-$7037 & $0.42\pm 0.03$ & $0.89\pm 0.03$ & $0.630\pm 0.011$ \\ 
19 & J0541$-$6800 & $0.08\pm 0.05$ & $0.30\pm 0.02$ & $0.667\pm 0.012$ \\
20 & J0541$-$6815 & $0.21\pm 0.03$ & $0.27\pm 0.02$ & $0.643\pm 0.012$ \\
&&&&\\
21 & J0547$-$7207 & $0.03\pm 0.02$ & $0.09\pm 0.01$ & $0.655\pm 0.014$ \\
22 & J0551$-$6916$^{\star}$ & $0.06\pm 0.04$ & $0.046\pm 0.004$ & $0.607\pm 0.016$ \\
23 & J0551$-$6843$^{\star}$ & $0.11\pm 0.04$ & $0.065\pm 0.005$ & $0.623\pm 0.014$ \\
24 & J0552$-$6850 & $0.22\pm 0.05$ & $0.25\pm 0.02$ & $0.706\pm 0.013$ \\
25 & J0557$-$6944 & $0.49\pm 0.05$ & $0.26\pm 0.03$ & $0.684\pm 0.014$  \\
&&&&\\
26 & J0559$-$6920 & $0.22\pm 0.03$ & $0.03\pm 0.02$ & $0.647\pm 0.014$ \\
27 & J0602$-$6830 & $0.07\pm 0.04$ & $0.10\pm 0.01$ & $0.634\pm 0.014$ \\\hline
\multicolumn{5}{c}{BL Lac type blazar candidates}\\\hline
1 & J0039$-$7356 & $0.44\pm 0.03$ & $0.96\pm 0.02$ & $0.660\pm 0.010$ \\
2 & J0111$-$7302$^{\star}$ & $0.35\pm 0.04$ & $0.73\pm 0.03$ & $0.680\pm 0.012$ \\
3 & J0123$-$7236 & --- & $0.95\pm 0.03$ & $0.681\pm 0.012$ \\
4 & J0439$-$6832 & $0.60\pm 0.06$ & $0.83\pm 0.03$ & $0.658\pm 0.014$ \\
5 & J0441$-$6945 & $0.45\pm 0.03$ & $0.80\pm 0.04$ & $0.660\pm 0.014$ \\
&&&&\\
6 & J0444$-$6729 & $0.21\pm 0.05$ & $0.51\pm 0.04$ & $0.641\pm 0.019$ \\
7 & J0446$-$6718 & $0.58\pm 0.05$ & $0.94\pm 0.03$ & $0.655\pm 0.015$ \\
8 & J0453$-$6949 & $0.48\pm 0.04$ & $0.93\pm 0.03$ & $0.672\pm 0.012$ \\
9 & J0457$-$6920 & $0.26\pm 0.03$ & $0.66\pm 0.03$ & $0.651\pm 0.012$ \\
10 & J0501$-$6653$^{\star}$ & $0.29\pm 0.04$ & $0.39\pm 0.02$ & $0.643\pm 0.014$ \\
&&&&\\
11 & J0516$-$6803 & $0.62\pm 0.05$ & $0.99\pm 0.03$ & $0.670\pm 0.014$ \\
12 & J0518$-$6755$^{\star}$ & $0.18\pm 0.03$ & $0.26\pm 0.02$ & $0.665\pm 0.015$ \\
13 & J0521$-$6959 & $0.23\pm 0.04$ & $0.48\pm 0.03$ & $0.659\pm 0.016$ \\
14 & J0522$-$7135 & $0.39\pm 0.04$ & $0.88\pm 0.03$ & $0.657\pm 0.014$ \\
15 & J0538$-$7225 & $0.28\pm 0.04$ & $0.32\pm 0.02$ & $0.659\pm 0.014$ \\
&&&&\\
16 & J0545$-$6846 & $0.58\pm 0.05$ & $0.87\pm 0.04$ & $0.658\pm 0.019$ \\
17 & J0553$-$6845 & $0.36\pm 0.04$ & $0.58\pm 0.02$ & $0.634\pm 0.013$
\enddata 
\tablecomments{$^\star$Sources considered as secure blazar candidates by \citet{Zywu18}. \\Columns: (1) number of the source; (2) source designation; (3) Hurst exponent; (4) Abbe value; (5) ratio of turning points. }
\end{deluxetable*}

\begin{figure}
\centering
\includegraphics[width=\columnwidth]{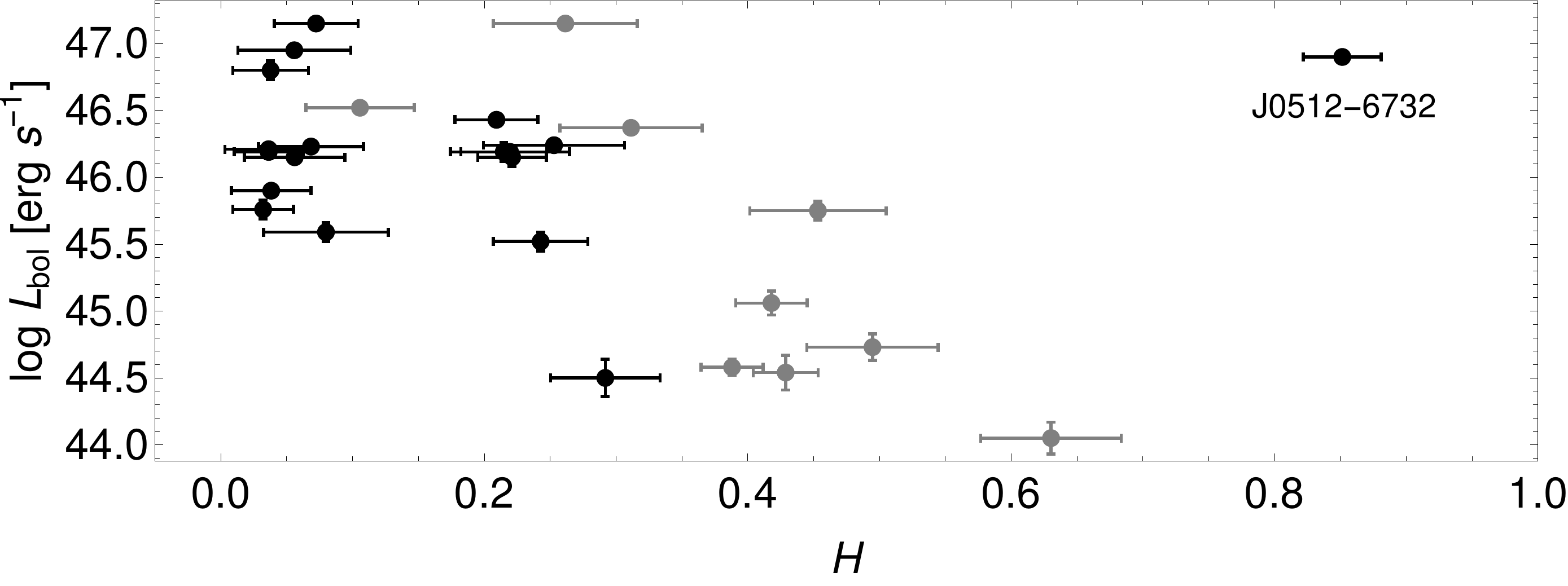}
\caption{Correlation between $H$ and $\log L_{\rm bol}$ for FSRQ blazar type candidates. The black symbols correspond to the 18 objects with PSDs given by model B (including J0527$-$7036, see Table~\ref{results}), while the gray symbols denote the remaining ones.}
\label{Lbolo_H}
\end{figure}

We note an interesting correlation between $H$ and the bolometric luminosities for FSRQ type candidates (see Table~\ref{blazarlist3} and Sect.~\ref{sect_mass2}), displayed in Fig.~\ref{Lbolo_H}. The Pearson coefficient is $r=-0.41$; after discarding J0512$-$6732, a bright outlier with $H=0.85\pm 0.03$, we obtain $r=-0.70$. While potentially this could link the persistence properties of an LC (via $H$) and physical processes governing the radiative output (via $L_{\rm bol}$), we propose a more straightforward explanation: in our FSRQ sample, it turns out the dimmer the object, the lower the $L_{\rm bol}$ ($r=-0.89$), and so, as we argued in Sect.~\ref{results_lsp}, the more the LC is consistent with white noise (Poisson noise level dominates the PSD). Indeed, most objects with PSDs given by model A lie roughly around $H\approx 0.5$. On the other hand, J0512$-$6732 does not follow this scheme, as it is one of the brightest FSRQ candidates in our sample and yields a remarkably high value of $H$. Recall that this source's PSD is better described by model B, with an exceptionally short $T_{\rm break} = 67\pm 10\,{\rm d}$. Finally, cases with $H\lesssim 0.2$ are dubious due to the discontinuity of $H$ (see Fig.~\ref{fig_Hurst_disc}), so they might as well yield high values of $H$. It is therefore unclear whether this one outlier is a statistical fluctuation, or a hint of a subpopulation of bright, long-term memory blazars.

\subsection{The $\mathcal{A}-\mathcal{T}$ plane}
\label{results_AT}

\begin{figure}
\centering
\includegraphics[width=\columnwidth]{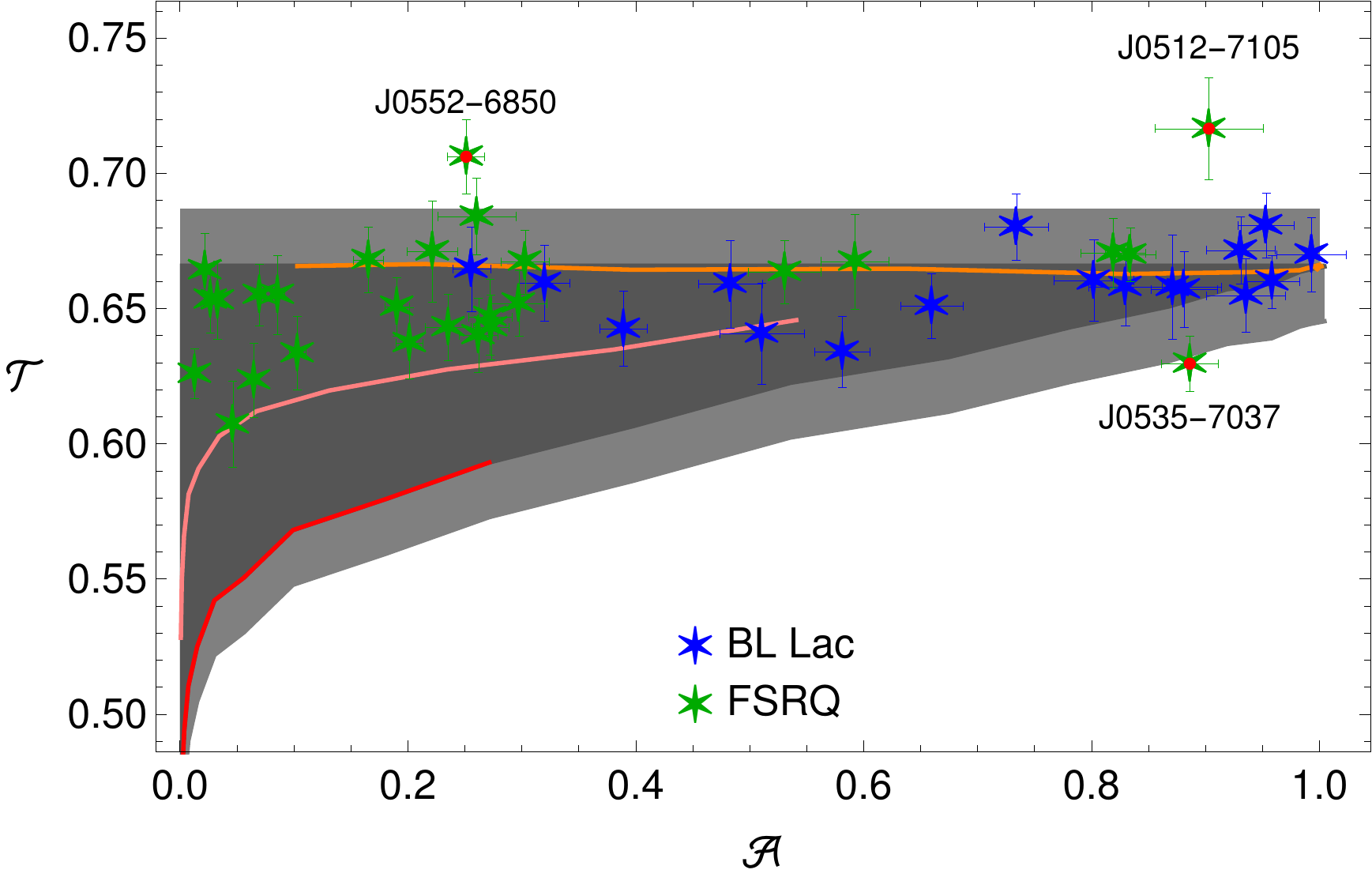}
\caption{Locations in the $\mathcal{A-T}$ plane of the blazar candidates. The dark gray area is the region between the pure PL line (red curve from Fig.~\ref{fig_AT}) and $\mathcal{T}=2/3$, and the light gray regions correspond to the error bars of the simulations. The red, pink and orange lines correspond to Fig.~\ref{fig_AT}, but only the part $\beta\in[1,2]$ is displayed herein (see Sect.~\ref{meth_AT}). Two FSRQs lie above the region admitted by model A noises, and one is located marginally below.}
\label{AT_results}
\end{figure}

The locations of FSRQs and BL Lacs in the $\mathcal{A}-\mathcal{T}$ plane are gathered in Table~\ref{blazarlist2}, and displayed in Fig.~\ref{AT_results}. The uncertainties are computed by bootstrapping. Most objects fall in the region occupied by PL plus Poisson noise type processes, with various Poisson noise levels. Three FSRQs are interestingly placed: J0535$-$7037 marginally below the pure PL line, while J0512$-$7105 and J0552$-$6850 above the limiting $\mathcal{T}=2/3$ line. The LSP implies that J0535$-$7037 has a PL PSD with $\beta\approx 1$, i.e. pink noise. J0512$-$7105 has a flat PSD, but yields $H>0.5$. The PSD of J0552$-$6850, however, shows a clear flattening on time scales greater than 1200 days, and has $H<0.5$. They are also distant in the $\mathcal{A}-\mathcal{T}$ plane, with $\mathcal{A} = 0.90$ and $\mathcal{A} = 0.25$, respectively.

One thing to bear in mind is that both $\mathcal{A}$ and $\mathcal{T}$ are order statistics, i.e. they are insensitive to the spacing between consecutive data points. One way of justifying the usage of the $\mathcal{A}-\mathcal{T}$ plane in case of irregularly sampled time series is by noting that any LC comes from sampling a continuous process, hence a continuum of data between any two observations is missing. In spite of this fact, it can be generally expected that the true characteristics of an analyzed system are properly captured by the observations, and the interrelation between available data points catches the overall behavior of the system.

Among our 44 blazar candidates, 41 are located in the region of the $\mathcal{A}-\mathcal{T}$ plane occupied by PL plus Poisson noise processes; one FSRQ is located marginally below ({\it less noisy} than white noise), and two FSRQ candidates are above the line $\mathcal{T}=2/3$ ({\it more noisy} than white noise). While not of direct interpretation herein, these two objects clearly stand out in this context. Moreover, BL Lac candidates are clearly characterized by higher $\mathcal{A}$ values than FSRQ ones (means of $0.71\pm 0.06$ and $0.29\pm 0.05$, respectively). This, as discussed in Sections~\ref{results_lsp} and \ref{results_hurst}, is consistent with dimmer objects being more noise polluted. Recent developments of the $\mathcal{A}-\mathcal{T}$ methodology \citep{zunino17,zhao18} prove it to be a useful tool in time series analysis.

\section{Discussion} \label{discussion}

\subsection{Physical processes}

The optical emission of blazars is expected to be predominantly due to the synchrotron radiation of highly relativistic electrons and positrons accelerated in-situ within the blazar emission zone. In the majority of theoretical models and scenarios proposed to account for the PL shape of blazars' PSDs, it is assumed that the energy density of the synchrotron-emitting electrons fluctuates about a mean value, with the distribution which is also a PL in the frequency domain
\citep[e.g.,][]{Mars2014}. Such fluctuations could be related to the dominant electron acceleration process at work, e.g. fluctuations in the bulk Lorentz factor at the base of a jet, which determine properties of the internal shocks developing further along the outflow and dissipating the jet bulk kinetic energy into the internal energy of the jet electrons \citep[see, e.g.,][]{Malz13,Malz14}. I.e., one \emph{assumes} a given PL form of the variability power spectrum of electron fluctuations, to match the observed PSD in a given range of the electromagnetic spectrum. As discussed by \citet{finke14,finke15}, for the electron variability power spectrum $\sim 1/f^{\beta}$, the index of
the corresponding PSD of the synchrotron emission is also $\beta$ on time scales longer than the characteristic cooling time scale of the emitting electrons, but $\beta+2$ on the time scales shorter than that.

This situation is analogous to the case of disk dominated systems, where perturbations in the local disk parameters (e.g. magnetic field), leading to the enhanced energy dissipation, shape the observed PSD of the thermal disk emission. In particular, \citet{kelly09,kelly11} discussed how uncorrelated (Gaussian) fluctuations within the disk may result in the observed PSD of the $\sim 1/f^0$ form on time scales longer than the characteristic relaxation (thermal) time scale in the system, and $\sim 1/f^2$ on the time scales shorter than this.

Based on the above reasoning and scenarios, for the sources most likely dominated by the disk emission, i.e. FSRQs and FSRQ candidates from our sample, we propose a possible interpretation for the breaks and high $\beta_2$ in the obtained PSDs by connecting them to the dynamical change of the accreting matter’s orbits at the inner edge of the disk, related with the thermal time scale $t_{\rm th}$. We explore the implications of such association in the subsequent sections.

\subsection{Overview of the optical PSDs of AGNs}

\citet{chatt08} analyzed an optical (R filter) LC of the FSRQ 3C 279, spanning 11 years. They found its PSD to be well described by a PL with $\beta=1.7$. Four Kepler AGNs observed over 2--4 quarters exhibited steep PSDs, with $\beta\approx 3$, within time scales of 1--100 days \citep{Mush11}. On the other hand, four other radio-loud AGNs, including three FSRQs, observed over 8--11 quarters, were characterized by $\beta\in (1,2)$ \citep{wehrle13,reval14}. \citet{simm16} examined $\approx 90$ AGNs from the Pan-STARRS1 XMM-COSMOS survey, and obtained in most cases good fits with a broken PL, with break time scales 100--300 days, a low-frequency PL index $\beta_1\in (0,2)$, and a high-frequency index $\beta_2\in (2,4)$. Recently, \citet{aranzana18} examined short-term variability of 252 Kepler AGNs, resulting in $\beta\in (1, 3.5)$, with a mode at 2.4, although the displayed PSDs clearly exhibit flattening at time scales $\gtrsim 10^{5-5.5}\,{\rm s}$. \citet{smith18} examined 21 Kepler AGNs, spanning  3--14 quarters, and in six cases found breaks within $\sim 10-50$ days in their PSDs. In particular, a mild correlation between $M_{\rm BH}$ and $T_{\rm break}$ was observed, contrary to \citet{simm16}. Finally, \citet{caplar19} investigated the PSDs of $\approx 2200$ AGNs from the Palomar Transient Factory survey, and obtained $\beta\in (1.5, 4)$.

\citet{Kast11} used three methods to investigate the long-term LC of a BL Lac object, PKS 2155$-$304, spanning the years 1934--2010, i.e. LSP, the structure function (SF), and the multiple fragments variance function (MFVF). Interestingly, they found $\beta=5.0^{+1.7}_{-3.0}$ using LSP, $\beta=1.6^{+0.4}_{-0.2}$ with SF, and $\beta=1.8^{+0.1}_{-0.2}$ with MFVF. All three methods give comparable results within errors with a break to an assumed white noise at time scales $\gtrsim 1000$ days. For six blazars (five FSRQs and one BL Lac), \citet{chatt12} obtained $\beta\approx 2$, with an exception of the FSRQ PKS 1510$-$089, which yielded a much flatter PSD with $\beta=0.6$. An R filter LC of the BL Lac PKS 0735+178 exhibits a purely red noise PSD, i.e. with $\beta=2$ \citep{goyal17}. Similarly, R filter LCs of 29 BL Lacs and two FSRQs, spanning $\approx$10 years, were analyzed by \citet{nilss18}, who obtained $\beta\in (1,2)$.

Overall, it appears that AGNs are either characterized by PSDs in the form of model A, with $\beta\in(1,3)$, occasionally with a flatter PSD, or exhibit breaks (model B) on time scales 100--1000 days, with a steeper PL component at high frequencies. The blazar candidates examined herein fall into those two categories: objects well described by model A yield $\beta\in(1,2)$, while those better characterized by model B exhibit break time scales within roughly 100--400 days, low-frequency PL index $\beta_1\lesssim 1$, and a steep PL component at higher frequencies, $\beta_2\in(3,7)$, reaching as high as 9 (see Table~\ref{blazarlist}). Such steep PSDs effectively imply no variability on the associated time scales, as the power drops drastically from the conventional PL at lower frequencies to the Poisson noise level. This means there is a sharp cutoff at $T_{\rm break}$ below which variability on short time scales is wiped out (excluding the region of Poisson noise domination). Therefore, these faint, distant sources might constitute a different, peculiar class of blazars.

\subsection{Mass, spin, and viscosity estimates}
\label{sect_mass1}

In sources for which the observed optical emission is dominated by the radiative output of accretion disks rather than jets, as is in fact expected for FSRQ type objects, the break time scale $T_B$ can, in principle, be connected to the BH mass and spin. Such a break might appear when the matter inspiraling in the disk transitions from bound orbits to a free-fall occurring for distances smaller than the inner radius of the disk, and can explain the high $\beta_2$. Therefore, by assuming that the disk extends sharply to the innermost stable circular orbit (ISCO) located at radius $r=r_{\rm ISCO}$ \citep{Mohan14}:
\begin{equation}
\begin{split}
T_B &= 2\pi (r_{\rm ISCO}^{3/2}+a_\star)(1+z)\frac{GM_{\rm BH}}{c^3} \\
&= 0.359\, m_9 (1+z) f(a_\star)\,{\rm day},
\end{split}
\label{eq14}
\end{equation}
where $m_9=M_{\rm BH}/(10^9 M_\odot)$, $a_\star=Jc/GM_{\rm BH}^2$ is the dimensionless spin, and $J$ the angular momentum of the BH. For a prograde rotation, $f(a_\star)$ is given by \citet{Bardeen72}:
\begin{subequations}
\begin{align}
& f(a_\star) = r_{\rm ISCO}^{3/2}+a_\star, \label{} \\
& r_{\rm ISCO} = 3 + Z_2-\sqrt{ \left( 3-Z_1 \right)\left( 3+Z_1+2Z_2 \right) }, \label{} \\
& Z_1 = 1+\left( 1-a_\star^2 \right)^{1/3} \left[ \left( 1+a_\star \right)^{1/3} + \left( 1-a_\star \right)^{1/3} \right], \label{} \\
& Z_2 = \sqrt{ 3a_\star^2 + Z_1^2 }, \label{}
\end{align}
\label{}%
\end{subequations}
We consider an accretion disk with a viscosity parameter $\alpha$ \citep{Shakura73,Novikov73,Page74}; then by associating the break time scale $T_{\rm break}$ from model B with the thermal time scale $t_{\rm th}$, we get \citep{czerny06,Lasota16}:
\begin{equation}
T_{\rm break} \sim t_{\rm th} \sim \alpha^{-1} t_{\rm K},
\end{equation}
where $t_K = 2\pi/\Omega$ is the orbital periodicity of a Keplerian motion on a circular orbit with radius $r$, with angular frequency $\Omega = (r^{3/2}+a_\star)^{-1}$ around the BH. Therefore, in Eq.~(\ref{eq14}), $T_B\sim t_{\rm K}\sim\alpha T_{\rm break}$.

With the values and uncertainties of $T_{\rm break}$ from Table~\ref{blazarlist} and redshifts from \citet{Zywu18}, we obtain a set of BH masses and spins satisfying Eq.~(\ref{eq14}). In Table~\ref{blazarlist3}, the 68\% confidence intervals\footnote{Corresponding to the uncertainties of $T_{\rm break}$.} for $\log M_{\rm BH}$ are given assuming $\alpha=0.1$ \citep{liu08}. The lower limits of the intervals are obtained for $T_{\rm break} - \Delta T_{\rm break}$ and $a_\star=0$; the upper limits are obtained for $T_{\rm break} + \Delta T_{\rm break}$ and $a_\star=0.998$, the maximal spin of an accreting BH \citep{Thorne74}. The value of $\alpha$ is accurate to a multiplicative factor of $\sim 3$ \citep{Grzedzielski17}, which can change $\log M_{\rm BH}$ by $\pm\log 3\approx \pm 0.48$. Moreover, the emission need not necessarily come from the ISCO, as, e.g., in a truncated disk model. By assuming, e.g., $r=2r_{\rm ISCO}$, the $\log M_{\rm BH}$ estimates are lowered by $0.45$ for $a_\star = 0$, and by $0.3$ for $a_\star = 0.998$. For $r=3r_{\rm ISCO}$, the respective factors are $0.7$ and $0.5$. Hints at the existence of truncated disks around supermassive BHs (SMBHs) were obtained for the radio galaxies: 3C 120 with a relatively low $\log M_{\rm BH} = 7.74$ \citep{cowp12,lohfink13}, and 4C+74.26 with $\log M_{\rm BH} = 9.6$ \citep{gofford15, Bhat18}, suggesting that the mass estimates from Table~\ref{blazarlist3} may be lowered by this account (but can be simultaneously increased if $\alpha>0.1$). On the other hand, our estimates are consistent with masses of bright Fermi blazars that are within $8\lesssim \log M_{\rm BH} \lesssim 10$ \citep{Ghis10a}, with FSRQs on average more massive than BL Lacs. A more recent sample of bright FSRQs only has a similar $\log M_{\rm BH}$ distribution \citep{Castignani13}. Moreover, some BH masses of FSRQs are known to attain high values, even up to $\log M_{\rm BH} = 10.6$ \citep{Ghis10b}. Therefore, our upper limits on the BH masses seem reasonable, and are consistent with the upper limits derived theoretically by \citet{inayoshi16} and \citet{king16}. They could be further constrained with dedicated observations.

\begin{deluxetable*}{clclc}
\tabletypesize{\footnotesize}
\tablecolumns{5}
\tablewidth{0pt}
\tablecaption{Bolometric luminosities and BH mass estimates of FSRQ type blazar candidates. \label{blazarlist3}}
\tablehead{
\multicolumn{1}{c}{Number} & \multicolumn{1}{c}{Object} & \colhead{$\log M_{\rm BH}^\dagger$} & \colhead{$\log L_{\rm bol}$} & \colhead{$\log M_{\rm BH}^\ddagger$} \\
 & & $[M_\odot]$ & \multicolumn{1}{c}{[${\rm erg}\,{\rm s}^{-1}$]} & $[M_\odot]$ \\
\tiny{(1)} & \multicolumn{1}{c}{\tiny{(2)}} & \tiny{(3)} & \multicolumn{1}{c}{\tiny{(4)}} & \tiny{(5)}}
\startdata
\multicolumn{5}{c}{FSRQ type blazar candidates}\\\hline
1 & J0054$-$7248 & --- & $44.54\pm 0.13$ & --- \\   
2 & J0114$-$7320$^{\star}$ & $(9.32,10.45)$ & $46.15\pm 0.05$ & $9.31\pm 0.31$ \\
3 & J0120$-$7334$^{\star}$ & $(9.06,10.14)$ & $46.80\pm 0.07^\diamond$ & $9.53\pm 0.34$ \\
4 & J0122$-$7152 & $(8.97,10.11)$ & $45.52\pm 0.07^\diamond$ & $8.86\pm 0.26$ \\
5 & J0442$-$6818$^{\star}$ & $(9.27,10.24)$ & $46.19\pm 0.05$ & $9.26\pm 0.30$ \\
&&&&\\
6 & J0445$-$6859 & --- & $46.37\pm 0.03$ & --- \\
7 & J0446$-$6758 & --- & $45.75\pm 0.07$ & --- \\
8 & J0455$-$6933 & $(9.20,10.19)$ & $46.24\pm 0.04$ & $9.29\pm 0.30$ \\
9 & J0459$-$6756 & $(9.01,10.18)$ & $46.19\pm 0.07^\diamond$ & $9.27\pm 0.31$ \\
10 & J0510$-$6941 & $(9.22,10.16)$ & $46.21\pm 0.03$ & $9.25\pm 0.30$ \\
&&&&\\
11 & J0512$-$7105 & --- & $44.05\pm 0.12$ & --- \\
12 & J0512$-$6732$^{\star}$ & $(8.49,9.40)$ & $46.90\pm 0.03$ & $9.33\pm 0.33$ \\
13 & J0515$-$6756 & --- & $44.58\pm 0.06$ & --- \\
14 & J0517$-$6759 & $(9.16,10.25)$ & $44.50\pm 0.14$ & $8.39\pm 0.22$ \\
15 & J0527$-$7036 & $(8.18,9.73)$ & $45.90\pm 0.04$ & $8.79\pm 0.30$ \\
&&&&\\ 
16 & J0528$-$6836 & --- & $47.15\pm 0.02$ & --- \\
17 & J0532$-$6931 & $(8.76,9.90)$ & $47.15\pm 0.02$ & $9.56\pm 0.36$ \\
18 & J0535$-$7037 & --- & $45.06\pm 0.09$ & --- \\
19 & J0541$-$6800 & $(9.16,10.47)$ & $45.59\pm 0.07^\diamond$ & $9.04\pm 0.29$ \\
20 & J0541$-$6815 & $(9.03,9.98)$ & $46.43\pm 0.03$ & $9.31\pm 0.31$ \\
&&&&\\
21 & J0547$-$7207 & $(9.28,10.40)$ & $45.76\pm 0.07^\diamond$ & $9.09\pm 0.28$ \\
22 & J0551$-$6916$^{\star}$ & $(9.01,10.00)$ & $46.95\pm 0.03$ & $9.60\pm 0.35$ \\
23 & J0551$-$6843$^{\star}$ & --- & $46.52\pm 0.04$ & --- \\
24 & J0552$-$6850 & $(9.74,10.84)$ & $46.19\pm 0.04$ & $9.59\pm 0.32$ \\
25 & J0557$-$6944 & --- & $44.73\pm 0.10$ & --- \\
&&&&\\
26 & J0559$-$6920 & $(9.02,10.15)$ & $46.15\pm 0.07$ & $9.24\pm 0.31$ \\
27 & J0602$-$6830 & $<10.79$ & $46.23\pm 0.03$ & $9.44\pm 0.38$
\enddata 
\tablecomments{$^\star$Sources considered as secure blazar candidates by \citet{Zywu18}.\\ $^\dagger$Assuming $\alpha=0.1$; the lower limit is for $T_{\rm break}-\Delta T_{\rm break}$ and $a_\star=0$; the upper limit is for $T_{\rm break}+\Delta T_{\rm break}$ and $a_\star=0.998$. See text for details. \\$^\diamond$Missing uncertainty of $I$ magnitude; the error of $\log L_{\rm bol}$ is estimated as the mean error of other objects. \\ $^\ddagger$ Estimates from the fundamental plane of AGN variability. \\ Columns: (1) number of the source; (2) source designation; (3) range of BH mass based on Eq.~(\ref{eq14}); (4) bolometric luminosity; (5) mass of BH based on Eq.~(\ref{eq17}).}
\end{deluxetable*}

Most SMBHs inhabiting radio quiet galaxies appear to have spins $a_\star\gtrsim 0.5$ \citep{McClintock11,Reynolds13,Reynolds14}, hence the masses are inclined to lie in the upper half of the presented intervals. Quasars are expected, on average, to exhibit accretion efficiency $\eta >0.1$ \citep{Soltan82}, corresponding to $a_\star>0.67$ \citep{sadowski11}. \citet{elvis02} argued that SMBHs should yield $\eta >0.15$, i.e. $a_\star>0.88$. Even though it is not clear what spins' range should be expected for radio loud galaxies, blazars in particular, it seems reasonable to expect the rotation rates to be high.

\begin{figure}
\centering
\includegraphics[width=\columnwidth]{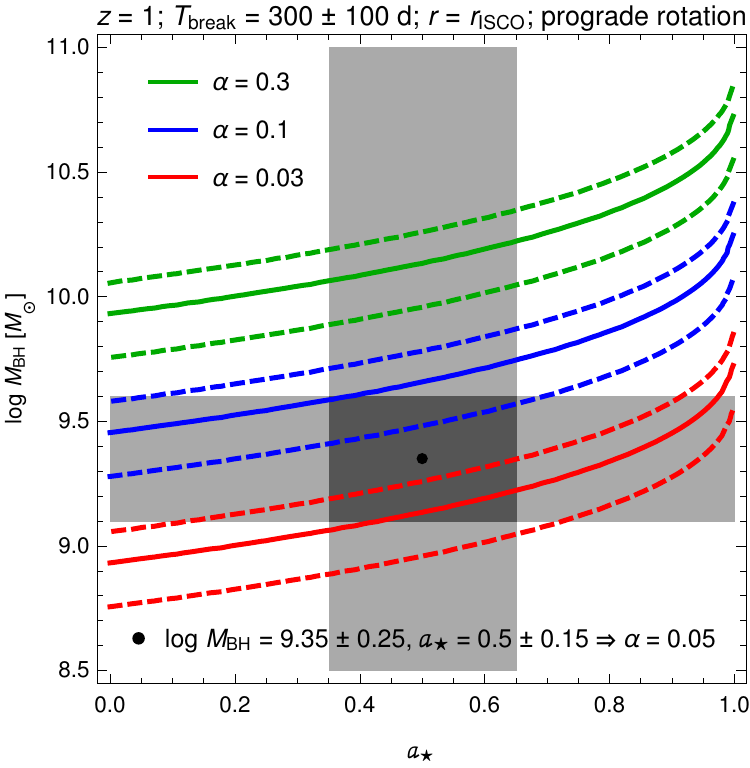}
\caption{BH mass and spin relation from Eq.~(\ref{eq14}) for prograde rotation, given $z$, $T_{\rm break}$, and three values of $\alpha$, and assuming emission comes from region close to ISCO. A hypothetical BH mass and spin, given in the bottom of the figure and highlighted by gray rectangles, indicates $\alpha = 0.05$. }
\label{BH_mass}
\end{figure}

In Fig.~\ref{BH_mass}, the effect of viscosity is presented for $\alpha = 0.03,\,0.1,\,0.3$, for some representative values of $z$ and $T_{\rm break}$. Overall, if the viscosity is not constrained tightly, $\log M_{\rm BH}$ is uncertain to an additive factor $\gtrsim 2$. On the other hand, if the BH mass and spin can be obtained with other methods, e.g. continuum-fitting, Fe ${\rm K}_{\alpha}$ line, or X-ray reflection for the spin \citep{McClintock11,Middleton16,kamm18}, and reverberation mapping \citep{Peterson14} for the mass, Eq.~(\ref{eq14}) can be used to estimate the viscosity parameter $\alpha$. In Fig.~\ref{BH_mass} such a hypothetical scenario is presented for $\log M_{\rm BH}$ and $a_\star$ with errors within the gray bands, with the black dot denoting the exact values, fixed in the simulation. It can be found that for $\alpha = 0.05$ one gets agreement between all relevant parameters.

Possibilities of retrograde rotation in AGNs were considered before \citep{Garofalo10}. We find, however, that then the dependence of $\log M_{\rm BH}$ on $a_\star$ is weakly negative, obviously coinciding for $a_\star=0$ with predictions from the prograde scenario. Whether the rotation is prograde or retrograde is another factor to take into account, although known BH spins suggest prograde rotation is more common.  

\subsection{Mass estimates from the fundamental plane of AGN variability}
\label{sect_mass2}

\citet{mchardy06} discovered a relation between the break time scale, BH mass, and bolometric luminosity to be:
\begin{equation}
\log \frac{T_{\rm break}}{1\,{\rm day}} = A \log \frac{M_{\rm BH}}{10^6 M_\odot} -B \log \frac{L_{\rm bol}}{10^{44}\,{\rm erg}\,{\rm s}^{-1}} + C,
\label{eq17}
\end{equation}
where $A=2.1\pm0.15$, $B=0.98\pm 0.15$, $C=-2.32\pm 0.2$.

\citet{marshall09} subsequently compared $T_{\rm break}$ from Eq.~(\ref{eq17}) with the estimate from the PSD, and obtained good agreement. However, Eq.~(\ref{eq17}) was obtained using X-ray data, hence the $T_{\rm break}$ therein refers to the X-ray PSD. Indeed, \citet{carini12} found that there is a discrepancy when the $T_{\rm break}$ is derived from the optical PSD. It is, however, still unclear whether ultimately the optical and X-ray characteristic time scales are the same or not \citep{smith18}.

Nevertheless, we employed Eq.~(\ref{eq17}) to estimate the BH masses. The bolometric luminosities were calculated according to \citet{Kozl15}\footnote{\url{http://www.astrouw.edu.pl/~simkoz/AGNcalc/}}, and are gathered in Table~\ref{blazarlist3}. For computing $L_{\rm bol}$, the latest cosmological parameters within a flat $\Lambda$CDM model \citep{Planck2018} are employed: $H_0=67.4\,{\rm km}\,{\rm s}^{-1}\,{\rm Mpc}^{-1}$, $\Omega_m=0.315$, $\Omega_\Lambda=0.685$. The $M_{\rm BH}$ estimates are mostly consistent with values obtained in Sect.~\ref{sect_mass1}, except for J0517$-$6759, for which we obtain a discrepancy of $\gtrsim 1$ order of magnitude. This can imply, assuming that Eq.~(\ref{eq17}) gives a proper mass, that the accretion disk is truncated, and/or is characterized by a low viscosity parameter.

\iffalse
Additionally, a relation between $T_{\rm break}$ and ${\rm H}_{\beta,{\rm FWHM}}$ was also found \citep{mchardy06}:
\begin{equation}
\log \frac{T_{\rm break}}{1\,{\rm day}} = 4.2 \log \frac{{\rm H}_{\beta,{\rm FWHM}}}{1\,{\rm km}\,{\rm s}^{-1}}-14.43
\label{eq18}
\end{equation}
for radio quiet objects. It was shown, using 3C 120, to not necessarily be applicable to radio loud galaxies, \citep{marshall09}. On the other hand, the results obtained for a radio loud galaxy 3C 390.3 \citep{gliozzi09} turned out to be consistent with values obtained from other methods. Therefore, for the sake of it, we show the resulting FWHM values in Fig.~\ref{fig_fwhm}.
\begin{figure}
\centering
\includegraphics[width=\columnwidth]{HbetaFWHM.pdf}
\caption{FWHM estimates for the FSRQs fitted with model B, obtained from Eq.~(\ref{eq18}). }
\label{fig_fwhm}
\end{figure}
\fi

\section{Summary} \label{conclusions}

For the 27 FSRQ and 17 BL Lac candidates, for which there are long, homogeneous LCs with multiple observations, an LSP was fitted with two models: PL plus Poisson noise, and SBPL plus Poisson noise (models A and B, respectively). We have also estimated the Hurst exponents, and used the recently developed $\mathcal{A}-\mathcal{T}$ plane in order to classify the LCs. The main conclusions are as follows:

\begin{enumerate}
\item 18 FSRQs in our sample yield PSDs consistent with model B. However, only four BL Lacs exhibit a detectable break in their PSDs. This might mean that the disk domination can manifest itself in the PSD via a break on the order of a few hundred days. On the contrary, in case of BL Lacs, lack of such a break might suggest jet domination. In this context, the three BL Lac objects described by model B are either not actually BL Lacs, or are peculiar BL Lacs with a significant radiative output coming from the accretion disk. 
\item Most of the secure blazar candidates (5/6 FSRQs and 2/3 BL Lacs) have PSDs best described by model B, with $T_{\rm break}$ at $200-300\,{\rm days}$; one FSRQ and one BL Lac are consistent with model A.
\item In case of objects exhibiting model B, the high frequency spectral index $\beta_2$ mostly lies in the range $3-7$. This steepness is intriguing: it can indicate a new class of AGNs in which the short term variability is effectively wiped out.
\item Two FSRQ and four BL Lac candidates were found to exhibit $H>0.5$, indicating long-term memory of the underlying governing process. This suggests that more complicated stochastic models need to be considered as a source for the observed variability.
\item We employed the recently developed $\mathcal{A}-\mathcal{T}$ plane in order to classify LCs, and identified two FSRQ type objects, J0512$-$7105 and J0552$-$6850, that are located in a region not available for PL types of PSD. While the first exhibits a flat PSD, the second yields a broken PL with $T_{\rm break}>1000\,{\rm days}$, hence both are exceptions in our sample.
\item Estimated BH masses of 18 FSRQs based on the $T_{\rm{break}}$ values, taking into account all possible BH spins, fall in the range $8.18\leq\log (M_{\rm BH}/M_\odot)\leq 10.84$.
\item Using bolometric luminosities and employing the fundamental plane of AGN variability as an independent estimate for the BH masses, we obtain the range $8.4\leq\log (M_{\rm BH}/M_\odot)\leq 9.6$, with a mean error of 0.3.
\end{enumerate}

\acknowledgments
The authors thank the OGLE group for the photometric data of the analyzed sources. N\.{Z} work was supported by the Polish National Science Center (NCN) through the PRELUDIUM grant DEC-2014/15/N/ST9/05171. MT acknowledges support by the NCN through the OPUS grant No. 2017/25/B/ST9/01208. The work of MB is supported through the South African Research Chair Initiative of the National Research Foundation\footnote{Any opinion, finding and conclusion or recommendation expressed in this material is that of the authors and the NRF does not accept any liability in this regard.} and the Department of Science and Technology of South Africa, under SARChI Chair grant No. 64789. {\L}S and VM acknowledge support by the NCN through grant No. 2016/22/E/ST9/00061.

\clearpage
\appendix

\section{PSD fits}
\label{appC}

\begin{figure*}
\centering
\includegraphics[height=\textheight]{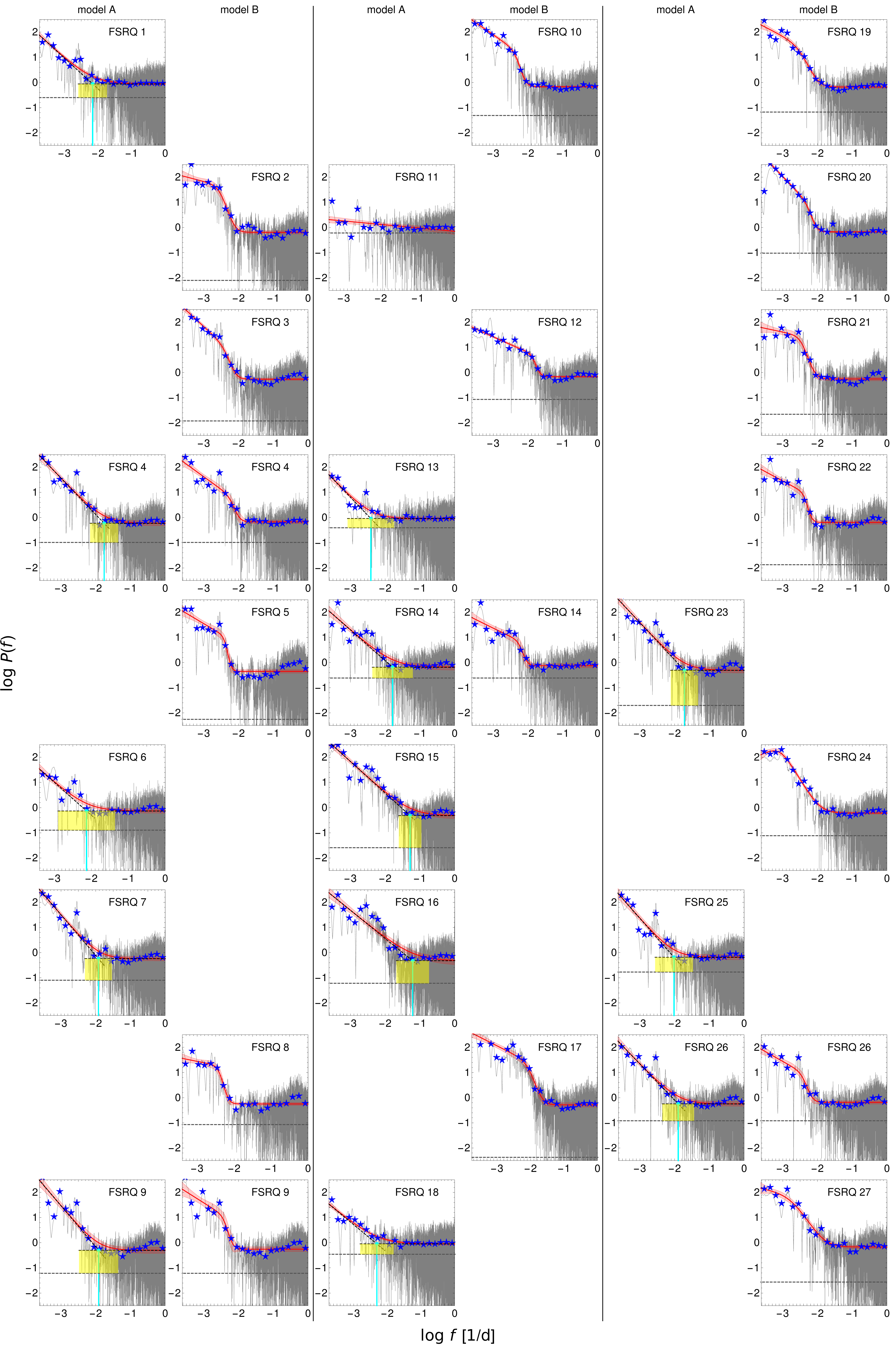}
\caption{Fits of models A and B, according to Table 1, of FSRQ blazar candidates. }
\label{fig_fits_FSRQ}
\end{figure*}

\begin{figure*}
\centering
\includegraphics[width=\columnwidth]{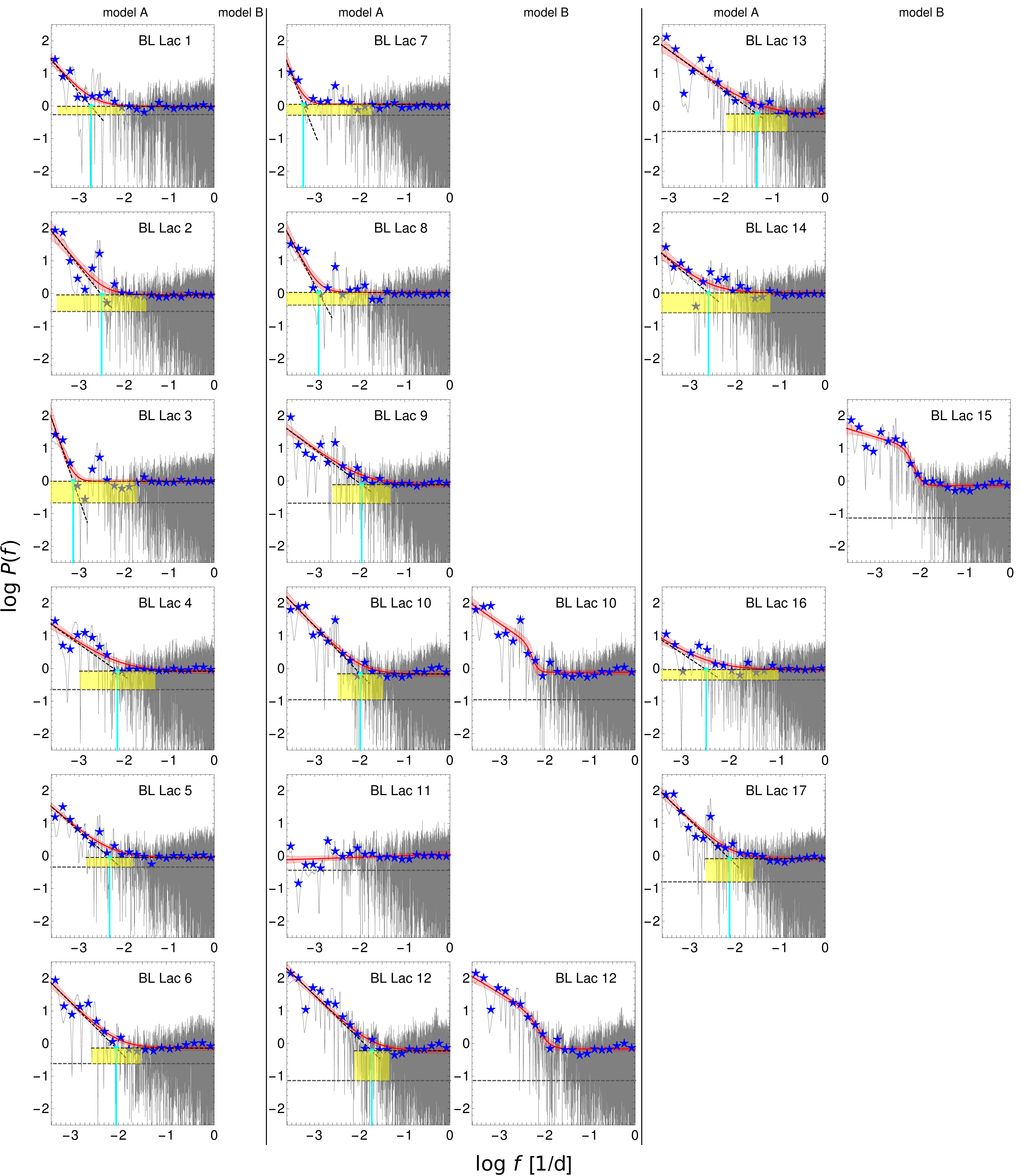}
\caption{Same as Fig.~\ref{fig_fits_FSRQ}, but for BL Lac blazar candidates.}
\label{fig_fits_BLLAC}
\end{figure*}

In Fig.~\ref{fig_fits_FSRQ} and \ref{fig_fits_BLLAC} we present fits of models A and B to the LSPs. In each panel, gray line is the raw LSP, and blue stars are the binned periodogram to which fitting was performed. The red solid line is the best fit, with the lighter red region around it marking the 68\% confidence interval. The black dashed lines are the PL component and Poisson noise level (in the model A column), whose intersection is marked with the cyan points. The vertical cyan line denotes the value of $f_0$. The width of the yellow rectangle denotes the standard error of $\log f_0$. The horizontal gray dashed line is the Poisson noise level inferred from data.

\section{LSP benchmark testing}
\label{appA}

%566,500,413,489,499,542,509

\begin{figure}
\centering
\includegraphics[width=0.67\textwidth]{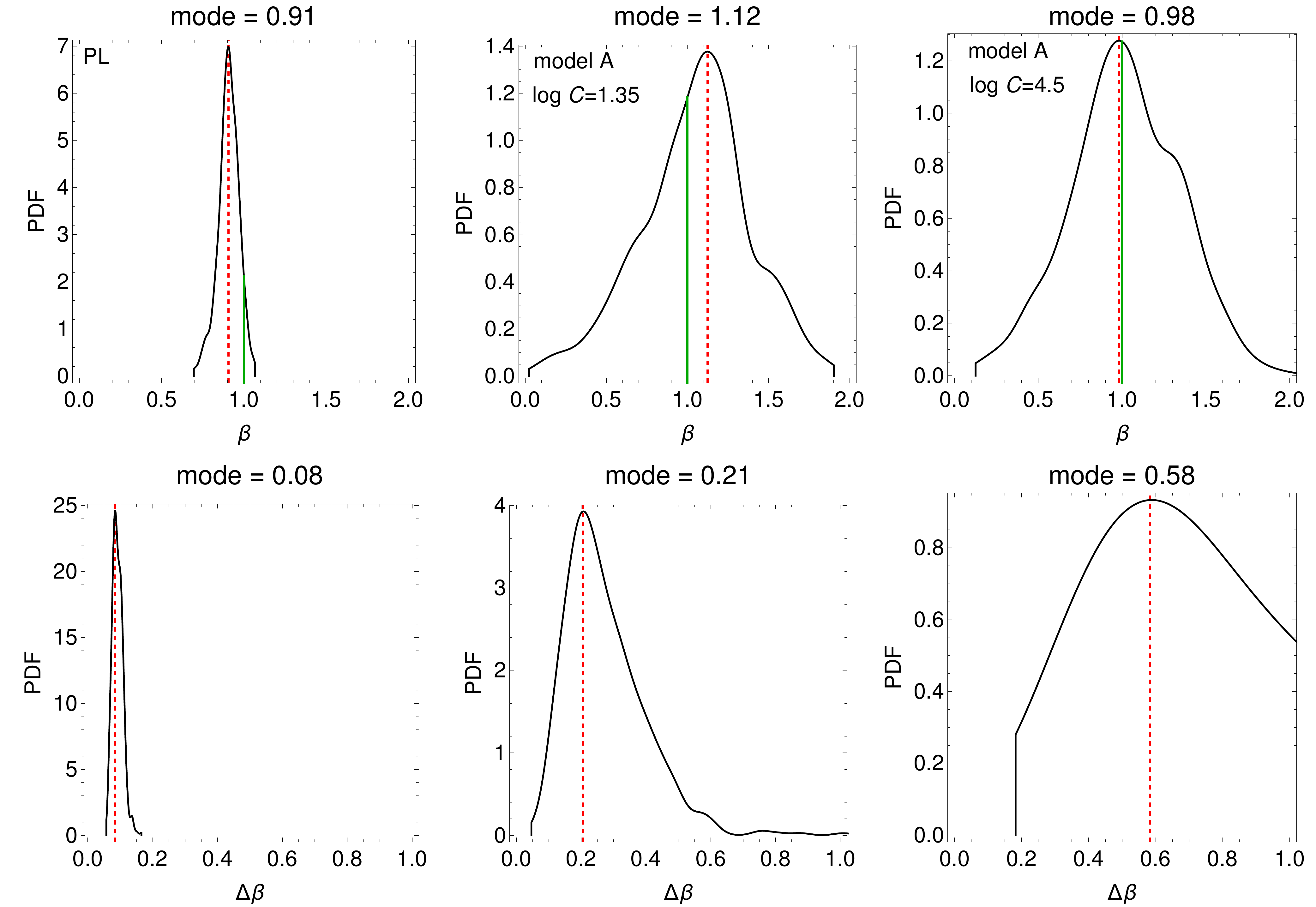}
\caption{The distributions of the PL index $\beta$ (upper row) and its error (bottom row) for the pure PL (left column), and model A with $\log C=1.35$ (middle column) and $\log C=4.5$ (right column), obtained from fitting to a LSP. The red dashed lines mark the modes of the distributions; the solid green line in the upper row denotes the input value $\beta=1$.}
\label{sh_1}
\end{figure}

\begin{figure}
\centering
\includegraphics[width=0.67\textwidth]{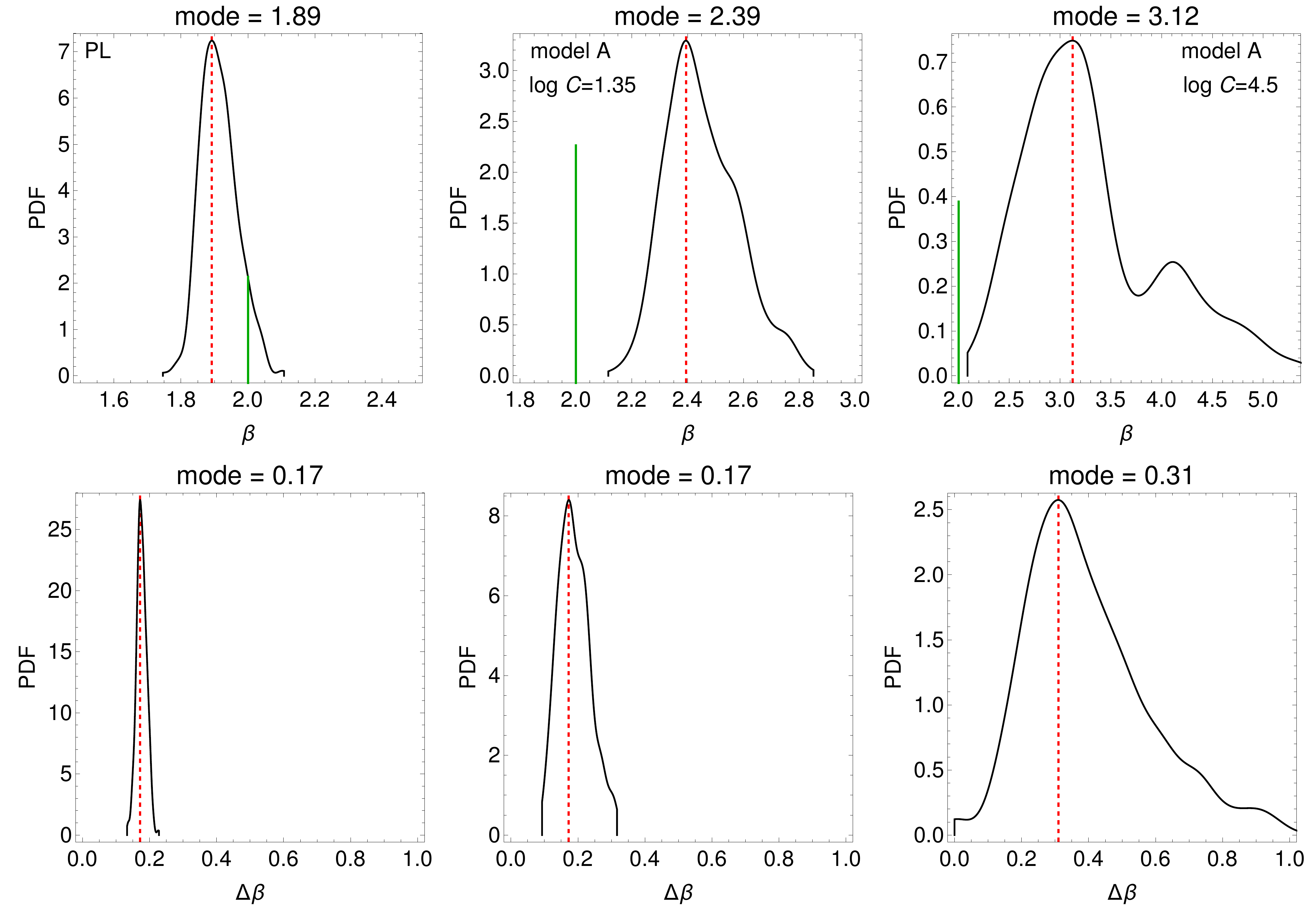}
\caption{Same as Fig.~\ref{sh_1}, but with input $\beta=2$.}
\label{sh_2}
\end{figure}

We generated in total $500$ LCs (that yielded fittable binned LSPs) of length $N=1024$ each (about the average number of points in OGLE LCs), from the pure PL and model A PSDs, with time step $\Delta t=1\,{\rm d}$, for combinations of values of $\beta=1$, $\beta=2$, $\log C=1.35$, and $\log C=4.5$. We then computed their LSPs, binned them, and fitted the PSD they were generated from, and recorded the obtained values of $\beta$ and their errors. The results are displayed in Fig.~\ref{sh_1} and \ref{sh_2}. The estimates obtained from a pure PL are close to the real values (slightly underestimated) and with reasonably small errors. When Poisson noise is introduced, the distributions of $\beta$ become wider, with much bigger errors returned. In particular, for the flatter PSDs with input $\beta=1$, the outputs span roughly from $\gtrsim 0$ to $\lesssim 2$. For input $\beta=2$ we observe a systematic overestimation of the PL index. For higher $C$, the output $\beta$ can exceed 5, and the error distribution is much wider than for lower $C$. 

\begin{figure}
\centering
\includegraphics[width=0.67\textwidth]{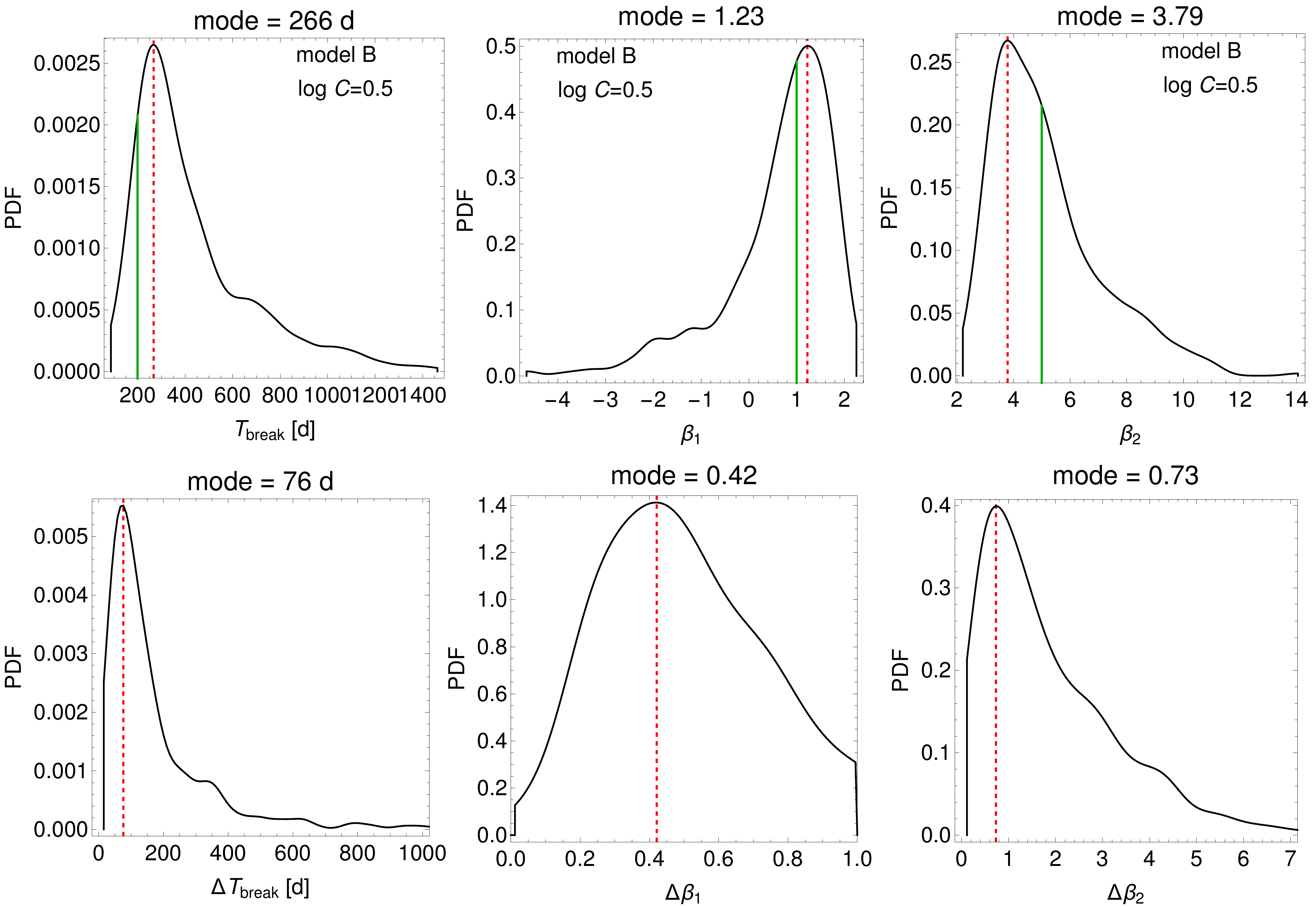}
\caption{Distributions of $T_{\rm break}$ (left column), $\beta_1$ (middle column) and $\beta_2$ (right column) for model B with $\log C = 0.5$. Note the scale of the horizontal axis on the rightmost lower panel. }
\label{sh_3}
\end{figure}

Next, model B was used to generate 500 LCs of length $N=2048$, with fittable binned LSPs. We chose the average values of parameters obtained from fitting the OGLE LCs: $\log C = 0.5$, $\log f_{\rm break} = -2.3$ (corresponding to $T_{\rm break}=200\,{\rm d}$), $\beta_1 = 1$ and $\beta_2 = 5$. The resulting distributions of $T_{\rm break}$, $\beta_1$ and $\beta_2$ are shown in Fig.~\ref{sh_3}. Displayed are only those fits that yielded $\beta_2 \neq 0$ within the errors --- for actual OGLE LCs we also considered such fits as unreliable, or simply consistent with model A. We find that the break time scale $T_{\rm break}$ is slightly overestimated on average, but individual values span a whole order of magnitude. The modes of indices $\beta_1$ and $\beta_2$ are both close to the input values, $\beta_1$ slightly overestimated, and $\beta_2$ underestimated by about 25\%. However, both indices have quite long and heavy tails, extending far from the input values. Therefore, while on average the input parameters are successfully recovered, individual fits can suffer from large, impossible to overcome biases.

\section{Impact of spacing on the fitting}
\label{appB}

In order to verify how the distribution of gaps in the LCs affects the model selection, we proceeded as follows. For each FSRQ, we generated 100 time series from model A, with $\beta=2$ and $\log C=2.5$, of total time coverage of the corresponding FSRQ, and imposed the same gaps present in the real data of that object. We therefore obtain time series with the same sampling as the respective source was observed with, and with a known underlying, true PSD. Next, models A and B were fitted and the better description was selected as described in Sect~\ref{sect3.1} and \ref{sect3.2}. Such a procedure was undertaken for the sampling of every FSRQ candidate, resulting in a total of 2700 selections; a summary is given in Table~\ref{tableApp}. In 18 cases, model A was (correctly) selected more often than model B. In the real data, model A is the better one in only 9 cases, i.e. model B was selected twice as often as should be expected if it was only due to statistical fluctuations. We also observed that the $\beta$ index was systematically underestimated, most often resulting in a value around 1.5 or lower.

The same procedure was applied to the sampling of BL Lac type candidates. In this case, model B was selected more often in only two instances, while we find model B to be a plausible description of the real data for four objects, i.e. also twice as often as should be detected if due only to statistical fluctuations.

\begin{deluxetable*}{cccccccccccccccccccccccccccc}
\tabletypesize{\footnotesize}
\tablecolumns{28}
\tablewidth{0pt}
\tablecaption{Percentage of model A being selected when spacings of the respective sources were imposed on simulated LCs. \label{tableApp}}
\startdata
\hline\hline
\multicolumn{28}{c}{FSRQ type blazar candidates}\\\hline
Number & 1 & 2 & 3 & 4 & 5 & 6 & 7 & 8 & 9 & 10  & 11 & 12 & 13 & 14 & 15 & 16 & 17 & 18 & 19 & 20 & 21 & 22 & 23 & 24 & 25 & 26 & 27\\
Percentage & 65 & 35 & 73 & 71 & 67 & 89 & 37 & 74 & 34 & 30 & 76 & 66 & 42 & 52 & 35 & 68 & 40 & 48 & 67 & 62 & 59 & 37 & 66 & 60 & 64 & 51 & 76\\\hline
\multicolumn{28}{c}{BL Lac type blazar candidates}\\\hline
Number & 1 & 2 & 3 & 4 & 5 & 6 & 7 & 8 & 9 & 10  & 11 & 12 & 13 & 14 & 15 & 16 & 17 & --- & --- & --- & --- & --- & --- & --- & --- & --- & ---\\
Percentage & 79 & 70 & 66 & 80 & 64 & 88 & 60 & 47 & 53 & 69 & 41 & 89 & 99 & 55 & 55 & 68 & 59 & --- & --- & --- & --- & --- & --- & --- & --- & --- & ---
\enddata 
\end{deluxetable*}

%\software{\textsc{Mathematica} \citep{Mathematica}, \textsc{R} \citep[][\url{http://www.R-project.org}]{RTeam}, \textsc{liftLRD} \citep[][\url{https://CRAN.R-project.org/package=liftLRD}]{knight17}}%, \textsc{CARMA} \citep[][\url{https://github.com/bckelly80/carma_pack}]{Kell14}}

\bibliography{LC_OGLE}

\begin{thebibliography}{}
\expandafter\ifx\csname natexlab\endcsname\relax\def\natexlab#1{#1}\fi

\bibitem[{{Akaike}(1974)}]{akaike74}
{Akaike}, H. 1974, IEEE Transactions on Automatic Control, 19, 716

\bibitem[{{Aleksi{\'c}} {et~al.}(2015){Aleksi{\'c}}, {Ansoldi}, {Antonelli},
  {Antoranz}, {Babic}, {Bangale}, {Barrio}, {Becerra Gonz{\'a}lez}, {Bednarek},
  {Bernardini}, {Biasuzzi}, {Biland}, {Blanch}, {Bonnefoy}, {Bonnoli},
  {Borracci}, {Bretz}, {Carmona}, {Carosi}, {Colin}, {Colombo}, {Contreras},
  {Cortina}, {Covino}, {Da Vela}, {Dazzi}, {De Angelis}, {De Caneva}, {De
  Lotto}, {de O{\~n}a Wilhelmi}, {Delgado Mendez}, {Di Pierro}, {Dominis
  Prester}, {Dorner}, {Doro}, {Einecke}, {Eisenacher}, {Elsaesser},
  {Fern{\'a}ndez-Barral}, {Fidalgo}, {Fonseca}, {Font}, {Frantzen}, {Fruck},
  {Galindo}, {Garc{\'\i}a L{\'o}pez}, {Garczarczyk}, {Garrido Terrats}, {Gaug},
  {Godinovi{\'c}}, {Gonz{\'a}lez Mu{\~n}oz}, {Gozzini}, {Hadasch}, {Hanabata},
  {Hayashida}, {Herrera}, {Hose}, {Hrupec}, {Idec}, {Kadenius}, {Kellermann},
  {Knoetig}, {Kodani}, {Konno}, {Krause}, {Kubo}, {Kushida}, {La Barbera},
  {Lelas}, {Lewandowska}, {Lindfors}, {Lombardi}, {Longo}, {L{\'o}pez},
  {L{\'o}pez-Coto}, {L{\'o}pez-Oramas}, {Lorenz}, {Lozano}, {Makariev},
  {Mallot}, {Maneva}, {Mannheim}, {Maraschi}, {Marcote}, {Mariotti},
  {Mart{\'\i}nez}, {Mazin}, {Menzel}, {Mirand a}, {Mirzoyan}, {Moralejo},
  {Munar-Adrover}, {Nakajima}, {Neustroev}, {Niedzwiecki}, {Nievas Rosillo},
  {Nilsson}, {Nishijima}, {Noda}, {Orito}, {Overkemping}, {Paiano},
  {Palatiello}, {Paneque}, {Paoletti}, {Paredes}, {Paredes-Fortuny}, {Persic},
  {Poutanen}, {Prada Moroni}, {Prandini}, {Puljak}, {Reinthal}, {Rhode},
  {Rib{\'o}}, {Rico}, {Rodriguez Garcia}, {Saito}, {Saito}, {Satalecka},
  {Scalzotto}, {Scapin}, {Schultz}, {Schweizer}, {Shore}, {Sillanp{\"a}{\"a}},
  {Sitarek}, {Snidaric}, {Sobczynska}, {Stamerra}, {Steinbring}, {Strzys},
  {Takalo}, {Takami}, {Tavecchio}, {Temnikov}, {Terzi{\'c}}, {Tescaro},
  {Teshima}, {Thaele}, {Torres}, {Toyama}, {Treves}, {Vogler}, {Will}, {Zanin},
  {Berger}, {Buson}, {D'Ammand o}, {Gasparrini}, {Hovatta}, {Max-Moerbeck},
  {Readhead}, \& {Richards}}]{Alek15}
{Aleksi{\'c}}, J., {Ansoldi}, S., {Antonelli}, L.~A., {et~al.} 2015, \mnras,
  451, 739

\bibitem[{{Aller} {et~al.}(2011){Aller}, {Aller}, \& {Hughes}}]{Alle11}
{Aller}, M.~F., {Aller}, H.~D., \& {Hughes}, P.~A. 2011, Journal of
  Astrophysics and Astronomy, 32, 5

\bibitem[{{Alston}(2019)}]{alston19b}
{Alston}, W.~N. 2019, \mnras, 485, 260

\bibitem[{{Alston} {et~al.}(2019){Alston}, {Fabian}, {Buisson}, {Kara},
  {Parker}, {Lohfink}, {Uttley}, {Wilkins}, {Pinto}, {De Marco}, {Cackett},
  {Middleton}, {Walton}, {Reynolds}, {Jiang}, {Gallo}, {Zogbhi}, {Miniutti},
  {Dovciak}, \& {Young}}]{alston19a}
{Alston}, W.~N., {Fabian}, A.~C., {Buisson}, D.~J.~K., {et~al.} 2019, \mnras,
  482, 2088

\bibitem[{{Angel} \& {Stockman}(1980)}]{Ange80}
{Angel}, J.~R.~P., \& {Stockman}, H.~S. 1980, \araa, 18, 321

\bibitem[{{Aranzana} {et~al.}(2018){Aranzana}, {K{\"o}rding}, {Uttley},
  {Scaringi}, \& {Bloemen}}]{aranzana18}
{Aranzana}, E., {K{\"o}rding}, E., {Uttley}, P., {Scaringi}, S., \& {Bloemen},
  S. 2018, \mnras, 476, 2501

\bibitem[{{Bachev} {et~al.}(2012){Bachev}, {Semkov}, {Strigachev}, {Gupta},
  {Gaur}, {Mihov}, {Boeva}, \& {Slavcheva-Mihova}}]{Bach12}
{Bachev}, R., {Semkov}, E., {Strigachev}, A., {et~al.} 2012, \mnras, 424, 2625

\bibitem[{{Bardeen} {et~al.}(1972){Bardeen}, {Press}, \&
  {Teukolsky}}]{Bardeen72}
{Bardeen}, J.~M., {Press}, W.~H., \& {Teukolsky}, S.~A. 1972, \apj, 178, 347

\bibitem[{{Bauer} {et~al.}(2009){Bauer}, {Baltay}, {Coppi}, {Donalek}, {Drake},
  {Djorgovski}, {Ellman}, {Glikman}, {Graham}, {Jerke}, {Mahabal},
  {Rabinowitz}, {Scalzo}, \& {Williams}}]{Baue09}
{Bauer}, A., {Baltay}, C., {Coppi}, P., {et~al.} 2009, \apj, 705, 46

\bibitem[{{Bhatta} {et~al.}(2018){Bhatta}, {Stawarz}, {Markowitz},
  {Balasubramaniam}, {Zola}, {Zdziarski}, {Jamrozy}, {Ostrowski}, {Kuzmicz},
  {Og{\l}oza}, {Dr{\'o}{\.z}d{\.z}}, {Siwak}, {Kozie{\l}-Wierzbowska},
  {Debski}, {Kundera}, {Stachowski}, {Machalski}, {Paliya}, \&
  {Caton}}]{Bhat18}
{Bhatta}, G., {Stawarz}, {\L}., {Markowitz}, A., {et~al.} 2018, \apj, 866, 132

\bibitem[{{Brockwell} \& {Davis}(1996)}]{brockwell96}
{Brockwell}, P.~J., \& {Davis}, R.~A. 1996, {Time Series: Theory and Methods,
  2nd ed.} (Springer-Verlag New York)

\bibitem[{Burnham \& Anderson(2004)}]{burnham04}
Burnham, K.~P., \& Anderson, D.~R. 2004, Sociological Methods \& Research, 33,
  261

\bibitem[{{Caplar} \& {Tacchella}(2019)}]{caplar19}
{Caplar}, N., \& {Tacchella}, S. 2019, arXiv e-prints

\bibitem[{{Carini} \& {Ryle}(2012)}]{carini12}
{Carini}, M.~T., \& {Ryle}, W.~T. 2012, \apj, 749, 70

\bibitem[{{Castignani} {et~al.}(2013){Castignani}, {Haardt}, {Lapi}, {De
  Zotti}, {Celotti}, \& {Danese}}]{Castignani13}
{Castignani}, G., {Haardt}, F., {Lapi}, A., {et~al.} 2013, \aap, 560, A28

\bibitem[{{Chatterjee} {et~al.}(2008){Chatterjee}, {Jorstad}, {Marscher}, {Oh},
  {McHardy}, {Aller}, {Aller}, {Balonek}, {Miller}, {Ryle}, {Tosti},
  {Kurtanidze}, {Nikolashvili}, {Larionov}, \& {Hagen-Thorn}}]{chatt08}
{Chatterjee}, R., {Jorstad}, S.~G., {Marscher}, A.~P., {et~al.} 2008, \apj,
  689, 79

\bibitem[{{Chatterjee} {et~al.}(2012){Chatterjee}, {Bailyn}, {Bonning},
  {Buxton}, {Coppi}, {Fossati}, {Isler}, {Maraschi}, \& {Urry}}]{chatt12}
{Chatterjee}, R., {Bailyn}, C.~D., {Bonning}, E.~W., {et~al.} 2012, \apj, 749,
  191

\bibitem[{{Cowperthwaite} \& {Reynolds}(2012)}]{cowp12}
{Cowperthwaite}, P.~S., \& {Reynolds}, C.~S. 2012, \apjl, 752, L21

\bibitem[{{Czerny}(2006)}]{czerny06}
{Czerny}, B. 2006, in Astronomical Society of the Pacific Conference Series,
  Vol. 360, AGN Variability from X-Rays to Radio Waves, ed. C.~M. {Gaskell},
  I.~M. {McHardy}, B.~M. {Peterson}, \& S.~G. {Sergeev}, 265

\bibitem[{{Elvis} {et~al.}(2002){Elvis}, {Risaliti}, \& {Zamorani}}]{elvis02}
{Elvis}, M., {Risaliti}, G., \& {Zamorani}, G. 2002, \apjl, 565, L75

\bibitem[{{Falomo} {et~al.}(2014){Falomo}, {Pian}, \& {Treves}}]{Falo14}
{Falomo}, R., {Pian}, E., \& {Treves}, A. 2014, \aapr, 22, 73

\bibitem[{{Feigelson} {et~al.}(2018){Feigelson}, {Babu}, \&
  {Caceres}}]{feigelson18}
{Feigelson}, E.~D., {Babu}, G.~J., \& {Caceres}, G.~A. 2018, Frontiers in
  Physics, 6, 80

\bibitem[{{Finke} \& {Becker}(2014)}]{finke14}
{Finke}, J.~D., \& {Becker}, P.~A. 2014, \apj, 791, 21

\bibitem[{{Finke} \& {Becker}(2015)}]{finke15}
---. 2015, \apj, 809, 85

\bibitem[{{Garofalo} {et~al.}(2010){Garofalo}, {Evans}, \&
  {Sambruna}}]{Garofalo10}
{Garofalo}, D., {Evans}, D.~A., \& {Sambruna}, R.~M. 2010, \mnras, 406

\bibitem[{{Gaur} {et~al.}(2010){Gaur}, {Gupta}, {Lachowicz}, \&
  {Wiita}}]{Gaur10}
{Gaur}, H., {Gupta}, A.~C., {Lachowicz}, P., \& {Wiita}, P.~J. 2010, \apj, 718,
  279

\bibitem[{{Gaur} {et~al.}(2012){Gaur}, {Gupta}, \& {Wiita}}]{Gaur12}
{Gaur}, H., {Gupta}, A.~C., \& {Wiita}, P.~J. 2012, \aj, 143, 23

\bibitem[{{Ghisellini} {et~al.}(2010{\natexlab{a}}){Ghisellini}, {Tavecchio},
  {Foschini}, {Ghirlanda}, {Maraschi}, \& {Celotti}}]{Ghis10a}
{Ghisellini}, G., {Tavecchio}, F., {Foschini}, L., {et~al.} 2010{\natexlab{a}},
  \mnras, 402, 497

\bibitem[{{Ghisellini} {et~al.}(2010{\natexlab{b}}){Ghisellini}, {Della Ceca},
  {Volonteri}, {Ghirlanda}, {Tavecchio}, {Foschini}, {Tagliaferri}, {Haardt},
  {Pareschi}, \& {Grindlay}}]{Ghis10b}
{Ghisellini}, G., {Della Ceca}, R., {Volonteri}, M., {et~al.}
  2010{\natexlab{b}}, \mnras, 405, 387

\bibitem[{{Gilfriche} {et~al.}(2018){Gilfriche}, {Deschodt-Arsac}, {Blons}, \&
  {Arsac}}]{gilfriche18}
{Gilfriche}, P., {Deschodt-Arsac}, V., {Blons}, E., \& {Arsac}, L.~M. 2018,
  Front. Physiol., 9, 293

\bibitem[{{Gofford} {et~al.}(2015){Gofford}, {Reeves}, {McLaughlin}, {Braito},
  {Turner}, {Tombesi}, \& {Cappi}}]{gofford15}
{Gofford}, J., {Reeves}, J.~N., {McLaughlin}, D.~E., {et~al.} 2015, \mnras,
  451, 4169

\bibitem[{{Goyal} {et~al.}(2017){Goyal}, {Stawarz}, {Ostrowski}, {Larionov},
  {Gopal-Krishna}, {Wiita}, {Joshi}, {Soida}, \& {Agudo}}]{goyal17}
{Goyal}, A., {Stawarz}, {\L}., {Ostrowski}, M., {et~al.} 2017, \apj, 837, 127

\bibitem[{{Grz{\k{e}}dzielski} {et~al.}(2017){Grz{\k{e}}dzielski}, {Janiuk},
  {Czerny}, \& {Wu}}]{Grzedzielski17}
{Grz{\k{e}}dzielski}, M., {Janiuk}, A., {Czerny}, B., \& {Wu}, Q. 2017, \aap,
  603, A110

\bibitem[{{Hartman} {et~al.}(1996){Hartman}, {Webb}, {Marscher}, {Travis},
  {Dermer}, {Aller}, {Aller}, {Balonek}, {Bennett}, {Bloom}, {Fujimoto},
  {Hermsen}, {Hughes}, {Jenkins}, {Kii}, {Kurfess}, {Makino}, {Mattox}, {von
  Montigny}, {Ohashi}, {Robson}, {Ryan}, {Sadun}, {Schoenfelder}, {Smith},
  {Teraesranta}, {Tornikoski}, \& {Turner}}]{Hart96}
{Hartman}, R.~C., {Webb}, J.~R., {Marscher}, A.~P., {et~al.} 1996, \apj, 461,
  698

\bibitem[{{Hurst}(1951)}]{hurst51}
{Hurst}, H.~E. 1951, Transactions of the American Society of Civil Engineers,
  116, 770

\bibitem[{Hurvich \& Tsai(1989)}]{hurvich89}
Hurvich, C.~M., \& Tsai, C.-L. 1989, Biometrika, 76, 297

\bibitem[{{Iler} {et~al.}(1997){Iler}, {Schachter}, \& {Birkinshaw}}]{Iler97}
{Iler}, A.~L., {Schachter}, J.~F., \& {Birkinshaw}, M. 1997, \apj, 486, 117

\bibitem[{{Inayoshi} \& {Haiman}(2016)}]{inayoshi16}
{Inayoshi}, K., \& {Haiman}, Z. 2016, \apj, 828, 110

\bibitem[{{Isobe} {et~al.}(2015){Isobe}, {Sato}, {Ueda}, {Hayashida},
  {Shidatsu}, {Kawamuro}, {Ueno}, {Sugizaki}, {Sugimoto}, {Mihara}, {Matsuoka},
  \& {Negoro}}]{isobe15}
{Isobe}, N., {Sato}, R., {Ueda}, Y., {et~al.} 2015, \apj, 798, 27

\bibitem[{{Kammoun} {et~al.}(2018){Kammoun}, {Nardini}, \& {Risaliti}}]{kamm18}
{Kammoun}, E.~S., {Nardini}, E., \& {Risaliti}, G. 2018, \aap, 614, A44

\bibitem[{{Kass} \& {Raftery}(1995)}]{kass95}
{Kass}, R.~E., \& {Raftery}, A.~E. 1995, J. Am. Stat. Assoc., 90, 773

\bibitem[{{Kastendieck} {et~al.}(2011){Kastendieck}, {Ashley}, \&
  {Horns}}]{Kast11}
{Kastendieck}, M.~A., {Ashley}, M.~C.~B., \& {Horns}, D. 2011, \aap, 531, A123

\bibitem[{{Katsev} \& {L'Heureux}(2003)}]{katsev03}
{Katsev}, S., \& {L'Heureux}, I. 2003, Computers and Geosciences, 29, 1085

\bibitem[{{Kellermann} {et~al.}(1989){Kellermann}, {Sramek}, {Schmidt},
  {Shaffer}, \& {Green}}]{Kell89}
{Kellermann}, K.~I., {Sramek}, R., {Schmidt}, M., {Shaffer}, D.~B., \& {Green},
  R. 1989, \aj, 98, 1195

\bibitem[{{Kelly} {et~al.}(2009){Kelly}, {Bechtold}, \&
  {Siemiginowska}}]{kelly09}
{Kelly}, B.~C., {Bechtold}, J., \& {Siemiginowska}, A. 2009, \apj, 698, 895

\bibitem[{{Kelly} {et~al.}(2011){Kelly}, {Sobolewska}, \&
  {Siemiginowska}}]{kelly11}
{Kelly}, B.~C., {Sobolewska}, M., \& {Siemiginowska}, A. 2011, \apj, 730, 52

\bibitem[{{Kendall} \& {Stuart}(1973)}]{kendall73}
{Kendall}, M., \& {Stuart}, A. 1973, {The advanced theory of statistics}
  (London: Griffin, 3rd ed.)

\bibitem[{{Kendall}(1971)}]{kendall1971}
{Kendall}, M.~G. 1971, Biometrika, 58, 369

\bibitem[{{King}(2016)}]{king16}
{King}, A. 2016, \mnras, 456, L109

\bibitem[{Knight {et~al.}(2017)Knight, Nason, \& Nunes}]{knight17}
Knight, M.~I., Nason, G.~P., \& Nunes, M.~A. 2017, Statistics and Computing,
  27, 1453

\bibitem[{{Koz{\l}owski}(2015)}]{Kozl15}
{Koz{\l}owski}, S. 2015, Acta Astron., 65, 251

\bibitem[{{Koz{\l}owski}(2016)}]{Kozl16}
---. 2016, \apj, 826, 118

\bibitem[{{Koz{\l}owski} \& {Kochanek}(2009)}]{Kozl09}
{Koz{\l}owski}, S., \& {Kochanek}, C.~S. 2009, \apj, 701, 508

\bibitem[{{Koz{\l}owski} {et~al.}(2010){Koz{\l}owski}, {Kochanek}, {Udalski},
  {Wyrzykowski}, {Soszy\'{n}ski}, {Szyma\'{n}ski}, {Kubiak}, {Pietrzy\'{n}ski},
  {Szewczyk}, {Ulaczyk}, {Poleski}, \& {OGLE Collaboration}}]{Kozl10}
{Koz{\l}owski}, S., {Kochanek}, C.~S., {Udalski}, A., {et~al.} 2010, \apj, 708,
  927

\bibitem[{{Koz{\l}owski} {et~al.}(2012){Koz{\l}owski}, {Kochanek}, {Jacyszyn},
  {Udalski}, {Szyma\'{n}ski}, {Poleski}, {Kubiak}, {Soszy\'{n}ski},
  {Pietrzy\'{n}ski}, {Wyrzykowski}, {Ulaczyk}, {Pietrukowicz}, \& {OGLE
  Collaboration}}]{Kozl12}
{Koz{\l}owski}, S., {Kochanek}, C.~S., {Jacyszyn}, A.~M., {et~al.} 2012, \apj,
  746, 27

\bibitem[{{Koz{\l}owski} {et~al.}(2013){Koz{\l}owski}, {Onken}, {Kochanek},
  {Udalski}, {Szyma\'{n}ski}, {Kubiak}, {Pietrzy\'{n}ski}, {Soszy\'{n}ski},
  {Wyrzykowski}, {Ulaczyk}, {Poleski}, {Pietrukowicz}, {Skowron}, {OGLE
  Collaboration}, {Meixner}, \& {Bonanos}}]{Kozl13}
{Koz{\l}owski}, S., {Onken}, C.~A., {Kochanek}, C.~S., {et~al.} 2013, \apj,
  775, 92

\bibitem[{{Lasota}(2016)}]{Lasota16}
{Lasota}, J.-P. 2016, in , 1

\bibitem[{{Liu} {et~al.}(2008){Liu}, {Bai}, {Zhao}, \& {Ma}}]{liu08}
{Liu}, H.~T., {Bai}, J.~M., {Zhao}, X.~H., \& {Ma}, L. 2008, \apj, 677, 884

\bibitem[{{Lohfink} {et~al.}(2013){Lohfink}, {Reynolds}, {Jorstad}, {Marscher},
  {Miller}, {Aller}, {Aller}, {Brenneman}, {Fabian}, {Miller}, {Mushotzky},
  {Nowak}, \& {Tombesi}}]{lohfink13}
{Lohfink}, A.~M., {Reynolds}, C.~S., {Jorstad}, S.~G., {et~al.} 2013, \apj,
  772, 83

\bibitem[{{Lomb}(1976)}]{lomb}
{Lomb}, N.~R. 1976, \apss, 39, 447

\bibitem[{{Malzac}(2013)}]{Malz13}
{Malzac}, J. 2013, \mnras, 429, L20

\bibitem[{{Malzac}(2014)}]{Malz14}
---. 2014, \mnras, 443, 299

\bibitem[{{Mandelbrot} \& {van Ness}(1968)}]{mandel68}
{Mandelbrot}, B.~B., \& {van Ness}, J.~W. 1968, SIAM Review, 10, 422

\bibitem[{{Marscher}(2014)}]{Mars2014}
{Marscher}, A.~P. 2014, \apj, 780, 87

\bibitem[{{Marshall} {et~al.}(2009){Marshall}, {Ryle}, {Miller}, {Marscher},
  {Jorstad}, {Chicka}, \& {McHardy}}]{marshall09}
{Marshall}, K., {Ryle}, W.~T., {Miller}, H.~R., {et~al.} 2009, \apj, 696, 601

\bibitem[{{McClintock} {et~al.}(2011){McClintock}, {Narayan}, {Davis}, {Gou},
  {Kulkarni}, {Orosz}, {Penna}, {Remillard}, \& {Steiner}}]{McClintock11}
{McClintock}, J.~E., {Narayan}, R., {Davis}, S.~W., {et~al.} 2011, Classical
  and Quantum Gravity, 28, 114009

\bibitem[{{McHardy} {et~al.}(2006){McHardy}, {Koerding}, {Knigge}, {Uttley}, \&
  {Fender}}]{mchardy06}
{McHardy}, I.~M., {Koerding}, E., {Knigge}, C., {Uttley}, P., \& {Fender},
  R.~P. 2006, \nat, 444, 730

\bibitem[{{McHardy} {et~al.}(2004){McHardy}, {Papadakis}, {Uttley}, {Page}, \&
  {Mason}}]{mchardy2004}
{McHardy}, I.~M., {Papadakis}, I.~E., {Uttley}, P., {Page}, M.~J., \& {Mason},
  K.~O. 2004, \mnras, 348, 783

\bibitem[{{Middleton}(2016)}]{Middleton16}
{Middleton}, M. 2016, in Astrophysics and Space Science Library, Vol. 440,
  Astrophysics of Black Holes: From Fundamental Aspects to Latest Developments,
  ed. C.~{Bambi}, 99

\bibitem[{{Mohan} \& {Mangalam}(2014)}]{Mohan14}
{Mohan}, P., \& {Mangalam}, A. 2014, \apj, 791, 74

\bibitem[{{Mowlavi}(2014)}]{mowlavi14}
{Mowlavi}, N. 2014, \aap, 568, A78

\bibitem[{{Murphy} {et~al.}(2010){Murphy}, {Sadler}, {Ekers}, {Massardi},
  {Hancock}, {Mahony}, {Ricci}, {Burke-Spolaor}, {Calabretta}, {Chhetri}, {de
  Zotti}, {Edwards}, {Ekers}, {Jackson}, {Kesteven}, {Lindley}, {Newton-McGee},
  {Phillips}, {Roberts}, {Sault}, {Staveley-Smith}, {Subrahmanyan}, {Walker},
  \& {Wilson}}]{Murp10}
{Murphy}, T., {Sadler}, E.~M., {Ekers}, R.~D., {et~al.} 2010, \mnras, 402, 2403

\bibitem[{{Mushotzky} {et~al.}(2011){Mushotzky}, {Edelson}, {Baumgartner}, \&
  {Gandhi}}]{Mush11}
{Mushotzky}, R.~F., {Edelson}, R., {Baumgartner}, W., \& {Gandhi}, P. 2011,
  \apjl, 743, L12

\bibitem[{{Nilsson} {et~al.}(2018){Nilsson}, {Lindfors}, {Takalo}, {Reinthal},
  {Berdyugin}, {Sillanp{\"a}{\"a}}, {Ciprini}, {Halkola}, {Hein{\"a}m{\"a}ki},
  {Hovatta}, {Kadenius}, {Nurmi}, {Ostorero}, {Pasanen}, {Rekola}, {Saarinen},
  {Sainio}, {Tuominen}, {Villforth}, {Vornanen}, \& {Zaprudin}}]{nilss18}
{Nilsson}, K., {Lindfors}, E., {Takalo}, L.~O., {et~al.} 2018, \aap, 620, A185

\bibitem[{{Nolan} {et~al.}(2012){Nolan}, {Abdo}, {Ackermann}, {Ajello},
  {Allafort}, {Antolini}, {Atwood}, {Axelsson}, {Baldini}, {Ballet}, \&
  et~al.}]{Nola12}
{Nolan}, P.~L., {Abdo}, A.~A., {Ackermann}, M., {et~al.} 2012, \apjs, 199, 31

\bibitem[{{Novikov} \& {Thorne}(1973)}]{Novikov73}
{Novikov}, I.~D., \& {Thorne}, K.~S. 1973, in Black Holes (Les Astres Occlus),
  ed. C.~{Dewitt} \& B.~S. {Dewitt}, 343--450

\bibitem[{{Page} \& {Thorne}(1974)}]{Page74}
{Page}, D.~N., \& {Thorne}, K.~S. 1974, \apj, 191, 499

\bibitem[{{Papadakis} \& {Lawrence}(1993)}]{papadakis93}
{Papadakis}, I.~E., \& {Lawrence}, A. 1993, \mnras, 261, 612

\bibitem[{Park \& Trippe(2014)}]{Park14}
Park, J.-H., \& Trippe, S. 2014, The Astrophysical Journal, 785, 76

\bibitem[{{Peterson}(2014)}]{Peterson14}
{Peterson}, B.~M. 2014, \ssr, 183, 253

\bibitem[{{Planck Collaboration} {et~al.}(2018){Planck Collaboration},
  {Aghanim}, {Akrami}, {Ashdown}, {Aumont}, {Baccigalupi}, {Ballardini},
  {Banday}, {Barreiro}, {Bartolo}, {Basak}, {Battye}, {Benabed}, {Bernard},
  {Bersanelli}, {Bielewicz}, {Bock}, {Bond}, {Borrill}, {Bouchet}, {Boulanger},
  {Bucher}, {Burigana}, {Butler}, {Calabrese}, {Cardoso}, {Carron},
  {Challinor}, {Chiang}, {Chluba}, {Colombo}, {Combet}, {Contreras}, {Crill},
  {Cuttaia}, {de Bernardis}, {de Zotti}, {Delabrouille}, {Delouis}, {Di
  Valentino}, {Diego}, {Dor{\'e}}, {Douspis}, {Ducout}, {Dupac}, {Dusini},
  {Efstathiou}, {Elsner}, {En{\ss}lin}, {Eriksen}, {Fantaye}, {Farhang},
  {Fergusson}, {Fernandez-Cobos}, {Finelli}, {Forastieri}, {Frailis},
  {Franceschi}, {Frolov}, {Galeotta}, {Galli}, {Ganga}, {G{\'e}nova-Santos},
  {Gerbino}, {Ghosh}, {Gonz{\'a}lez-Nuevo}, {G{\'o}rski}, {Gratton},
  {Gruppuso}, {Gudmundsson}, {Hamann}, {Hand ley}, {Herranz}, {Hivon}, {Huang},
  {Jaffe}, {Jones}, {Karakci}, {Keih{\"a}nen}, {Keskitalo}, {Kiiveri}, {Kim},
  {Kisner}, {Knox}, {Krachmalnicoff}, {Kunz}, {Kurki-Suonio}, {Lagache},
  {Lamarre}, {Lasenby}, {Lattanzi}, {Lawrence}, {Le Jeune}, {Lemos},
  {Lesgourgues}, {Levrier}, {Lewis}, {Liguori}, {Lilje}, {Lilley}, {Lindholm},
  {L{\'o}pez-Caniego}, {Lubin}, {Ma}, {Mac{\'\i}as-P{\'e}rez}, {Maggio},
  {Maino}, {Mandolesi}, {Mangilli}, {Marcos-Caballero}, {Maris}, {Martin},
  {Martinelli}, {Mart{\'\i}nez-Gonz{\'a}lez}, {Matarrese}, {Mauri}, {McEwen},
  {Meinhold}, {Melchiorri}, {Mennella}, {Migliaccio}, {Millea}, {Mitra},
  {Miville-Desch{\^e}nes}, {Molinari}, {Montier}, {Morgante}, {Moss}, {Natoli},
  {N{\o}rgaard-Nielsen}, {Pagano}, {Paoletti}, {Partridge}, {Patanchon},
  {Peiris}, {Perrotta}, {Pettorino}, {Piacentini}, {Polastri}, {Polenta},
  {Puget}, {Rachen}, {Reinecke}, {Remazeilles}, {Renzi}, {Rocha}, {Rosset},
  {Roudier}, {Rubi{\~n}o-Mart{\'\i}n}, {Ruiz-Granados}, {Salvati}, {Sandri},
  {Savelainen}, {Scott}, {Shellard}, {Sirignano}, {Sirri}, {Spencer},
  {Sunyaev}, {Suur-Uski}, {Tauber}, {Tavagnacco}, {Tenti}, {Toffolatti},
  {Tomasi}, {Trombetti}, {Valenziano}, {Valiviita}, {Van Tent}, {Vibert},
  {Vielva}, {Villa}, {Vittorio}, {Wand elt}, {Wehus}, {White}, {White},
  {Zacchei}, \& {Zonca}}]{Planck2018}
{Planck Collaboration}, {Aghanim}, N., {Akrami}, Y., {et~al.} 2018, arXiv
  e-prints

\bibitem[{{Press} \& {Rybicki}(1989)}]{press89}
{Press}, W.~H., \& {Rybicki}, G.~B. 1989, \apj, 338, 277

\bibitem[{{Rani} {et~al.}(2010){Rani}, {Gupta}, {Strigachev}, {Bachev},
  {Wiita}, {Semkov}, {Ovcharov}, {Mihov}, {Boeva}, {Peneva}, {Spassov},
  {Tsvetkova}, {Stoyanov}, \& {Valcheva}}]{Rani10}
{Rani}, B., {Gupta}, A.~C., {Strigachev}, A., {et~al.} 2010, \mnras, 404, 1992

\bibitem[{{Rani} {et~al.}(2013){Rani}, {Krichbaum}, {Fuhrmann}, {B{\"o}ttcher},
  {Lott}, {Aller}, {Aller}, {Angelakis}, {Bach}, {Bastieri}, {Falcone},
  {Fukazawa}, {Gabanyi}, {Gupta}, {Gurwell}, {Itoh}, {Kawabata}, {Krips},
  {L{\"a}hteenm{\"a}ki}, {Liu}, {Marchili}, {Max-Moerbeck}, {Nestoras},
  {Nieppola}, {Quintana-Lacaci}, {Readhead}, {Richards}, {Sasada}, {Sievers},
  {Sokolovsky}, {Stroh}, {Tammi}, {Tornikoski}, {Uemura}, {Ungerechts},
  {Urano}, \& {Zensus}}]{Rani13}
{Rani}, B., {Krichbaum}, T.~P., {Fuhrmann}, L., {et~al.} 2013, \aap, 552, A11

\bibitem[{{Revalski} {et~al.}(2014){Revalski}, {Nowak}, {Wiita}, {Wehrle}, \&
  {Unwin}}]{reval14}
{Revalski}, M., {Nowak}, D., {Wiita}, P.~J., {Wehrle}, A.~E., \& {Unwin}, S.~C.
  2014, \apj, 785, 60

\bibitem[{{Reynolds}(2013)}]{Reynolds13}
{Reynolds}, C.~S. 2013, Classical and Quantum Gravity, 30, 244004

\bibitem[{{Reynolds}(2014)}]{Reynolds14}
---. 2014, \ssr, 183, 277

\bibitem[{Richards {et~al.}(2014)Richards, Hovatta, Max-Moerbeck, Pavlidou,
  Pearson, \& Readhead}]{Rich14}
Richards, J.~L., Hovatta, T., Max-Moerbeck, W., {et~al.} 2014, Monthly Notices
  of the Royal Astronomical Society, 438, 3058

\bibitem[{{Ruan} {et~al.}(2012){Ruan}, {Anderson}, {MacLeod}, {Becker},
  {Burnett}, {Davenport}, {Ivezi{\'c}}, {Kochanek}, {Plotkin}, {Sesar}, \&
  {Stuart}}]{Ruan12}
{Ruan}, J.~J., {Anderson}, S.~F., {MacLeod}, C.~L., {et~al.} 2012, \apj, 760,
  51

\bibitem[{{Sagar} {et~al.}(2004){Sagar}, {Stalin}, {Gopal-Krishna}, \&
  {Wiita}}]{Saga04}
{Sagar}, R., {Stalin}, C.~S., {Gopal-Krishna}, \& {Wiita}, P.~J. 2004, \mnras,
  348, 176

\bibitem[{{Scargle}(1982)}]{scargle82}
{Scargle}, J.~D. 1982, \apj, 263, 835

\bibitem[{Schwarz(1978)}]{schwarz78}
Schwarz, G. 1978, Ann. Statist., 6, 461

\bibitem[{{Shakura} \& {Sunyaev}(1973)}]{Shakura73}
{Shakura}, N.~I., \& {Sunyaev}, R.~A. 1973, \aap, 24, 337

\bibitem[{{Simm} {et~al.}(2016){Simm}, {Salvato}, {Saglia}, {Ponti},
  {Lanzuisi}, {Trakhtenbrot}, {Nandra}, \& {Bender}}]{simm16}
{Simm}, T., {Salvato}, M., {Saglia}, R., {et~al.} 2016, \aap, 585

\bibitem[{{S\k adowski}(2011)}]{sadowski11}
{S\k adowski}, A. 2011, PhD thesis, Nicolaus Copernicus Astronomical Center,
  Polish Academy of Sciences

\bibitem[{{Smith} {et~al.}(2018){Smith}, {Mushotzky}, {Boyd}, {Malkan},
  {Howell}, \& {Gelino}}]{smith18}
{Smith}, K.~L., {Mushotzky}, R.~F., {Boyd}, P.~T., {et~al.} 2018, \apj, 857,
  141

\bibitem[{{Sobolewska} {et~al.}(2014){Sobolewska}, {Siemiginowska}, {Kelly}, \&
  {Nalewajko}}]{Sobo14}
{Sobolewska}, M.~A., {Siemiginowska}, A., {Kelly}, B.~C., \& {Nalewajko}, K.
  2014, \apj, 786

\bibitem[{{Soltan}(1982)}]{Soltan82}
{Soltan}, A. 1982, \mnras, 200, 115

\bibitem[{{Tarnopolski}(2015)}]{tarnopolski15c}
{Tarnopolski}, M. 2015, \mnras, 454, 1132

\bibitem[{{Tarnopolski}(2016)}]{tarnopolski16d}
---. 2016, Physica A Statistical Mechanics and its Applications, 461, 662

\bibitem[{{Thorne}(1974)}]{Thorne74}
{Thorne}, K.~S. 1974, \apj, 191, 507

\bibitem[{Tsai(2009)}]{tsai09}
Tsai, H. 2009, Bernoulli, 15, 178

\bibitem[{Tsai \& Chan(2005)}]{tsai05}
Tsai, H., \& Chan, K.~S. 2005, Journal of the Royal Statistical Society: Series
  B (Statistical Methodology), 67, 703

\bibitem[{{Udalski} {et~al.}(1997){Udalski}, {Kubiak}, \& {Szymanski}}]{Udal97}
{Udalski}, A., {Kubiak}, M., \& {Szymanski}, M. 1997, AcA, 47, 319

\bibitem[{{Udalski} {et~al.}(2015){Udalski}, {Szyma{\'n}ski}, \&
  {Szyma{\'n}ski}}]{Udal15}
{Udalski}, A., {Szyma{\'n}ski}, M.~K., \& {Szyma{\'n}ski}, G. 2015, AcA, 65, 1

\bibitem[{{Udalski} {et~al.}(2008{\natexlab{a}}){Udalski}, {Soszy\'{n}ski},
  {Szyma\'{n}ski}, {Kubiak}, {Pietrzy\'{n}ski}, {Wyrzykowski}, {Szewczyk},
  {Ulaczyk}, \& {Poleski}}]{Udal08a}
{Udalski}, A., {Soszy\'{n}ski}, I., {Szyma\'{n}ski}, M.~K., {et~al.}
  2008{\natexlab{a}}, AcA, 58, 89

\bibitem[{{Udalski} {et~al.}(2008{\natexlab{b}}){Udalski}, {Soszy{\'n}ski},
  {Szyma{\'n}ski}, {Kubiak}, {Pietrzy{\'n}ski}, {Wyrzykowski}, {Szewczyk},
  {Ulaczyk}, \& {Poleski}}]{Udal08b}
{Udalski}, A., {Soszy{\'n}ski}, I., {Szyma{\'n}ski}, M.~K., {et~al.}
  2008{\natexlab{b}}, AcA, 58, 329

\bibitem[{{Urry} \& {Padovani}(1995)}]{Urry95}
{Urry}, C.~M., \& {Padovani}, P. 1995, \pasp, 107, 803

\bibitem[{{VanderPlas}(2018)}]{vanderplas18}
{VanderPlas}, J.~T. 2018, \apjs, 236, 16

\bibitem[{Veitch \& Abry(1999)}]{veitch99}
Veitch, D., \& Abry, P. 1999, IEEE Transactions on Information Theory, 45, 878

\bibitem[{{Voges} {et~al.}(1999){Voges}, {Aschenbach}, {Boller},
  {Br{\"a}uninger}, {Briel}, {Burkert}, {Dennerl}, {Englhauser}, {Gruber},
  {Haberl}, {Hartner}, {Hasinger}, {K{\"u}rster}, {Pfeffermann}, {Pietsch},
  {Predehl}, {Rosso}, {Schmitt}, {Tr{\"u}mper}, \& {Zimmermann}}]{Voge99}
{Voges}, W., {Aschenbach}, B., {Boller}, T., {et~al.} 1999, \aap, 349, 389

\bibitem[{{von Neumann}(1941{\natexlab{a}})}]{neumann41a}
{von Neumann}, J. 1941{\natexlab{a}}, The Annals of Mathematical Statistics,
  12, 367

\bibitem[{{von Neumann}(1941{\natexlab{b}})}]{neumann41b}
---. 1941{\natexlab{b}}, The Annals of Mathematical Statistics, 12, 153

\bibitem[{{Wagner} \& {Witzel}(1995)}]{Wagn95}
{Wagner}, S.~J., \& {Witzel}, A. 1995, \araa, 33, 163

\bibitem[{{Wagner} {et~al.}(1996){Wagner}, {Witzel}, {Heidt}, {Krichbaum},
  {Qian}, {Quirrenbach}, {Wegner}, {Aller}, {Aller}, {Anton}, {Appenzeller},
  {Eckart}, {Kraus}, {Naundorf}, {Kneer}, {Steffen}, \& {Zensus}}]{Wagn96}
{Wagner}, S.~J., {Witzel}, A., {Heidt}, J., {et~al.} 1996, \aj, 111, 2187

\bibitem[{{Wehrle} {et~al.}(2013){Wehrle}, {Wiita}, {Unwin}, {Di Lorenzo},
  {Revalski}, {Silano}, \& {Sprague}}]{wehrle13}
{Wehrle}, A.~E., {Wiita}, P.~J., {Unwin}, S.~C., {et~al.} 2013, \apj, 773

\bibitem[{{Williams}(1941)}]{williams41}
{Williams}, J.~D. 1941, The Annals of Mathematical Statistics, 12, 239

\bibitem[{{Zhao} \& {Morales}(2018)}]{zhao18}
{Zhao}, Y., \& {Morales}, G.~J. 2018, \pre, 98, 022213

\bibitem[{{Zunino} {et~al.}(2017){Zunino}, {Olivares}, {Bariviera}, \&
  {Rosso}}]{zunino17}
{Zunino}, L., {Olivares}, F., {Bariviera}, A.~F., \& {Rosso}, O.~A. 2017,
  Physics Letters A, 381, 1021

\bibitem[{{{\.Z}ywucka} {et~al.}(2018){{\.Z}ywucka}, {Goyal}, {Jamrozy},
  {Stawarz}, {Ostrowski}, {Koz{\l}owski}, \& {Udalski}}]{Zywu18}
{{\.Z}ywucka}, N., {Goyal}, A., {Jamrozy}, M., {et~al.} 2018, \apj, 867, 131

\end{thebibliography}

\end{document}